\documentclass[11pt]{article}
\usepackage[authoryear]{natbib}
\usepackage{amssymb,amsmath,amsthm,bm}
\usepackage{fullpage}
\usepackage[edges]{forest} 
\usepackage{lineno}
\usepackage{hyperref}
\hypersetup{colorlinks=true,
	citecolor=blue,
	anchorcolor=red,
	urlcolor=black
}

\usepackage{epsfig}
\usepackage{enumerate}
\usepackage{xcolor}
\usepackage{extarrows}
\usepackage{graphicx}
\usepackage{multirow} 
\usepackage{float}
\usepackage{xr}

\bibpunct[, ]{[}{]}{,}{n}{,}{,}
\makeatletter
\def\NAT@def@citea{\def\@citea{\NAT@separator}}
\makeatother

\begin{document}

\title{A Regression Tree Method for Longitudinal and Clustered Data with Multivariate Responses}

\author{Wenbo Jing\footnote{Leonard N. Stern School of Business, New York University, e-mail: wj2093@stern.nyu.edu}\quad
	Jeffrey S. Simonoff\footnote{Leonard N. Stern School of Business, New York University}
}

\maketitle

\begin{abstract}
In this paper, we propose a tree-based method called Multivariate RE-EM tree, which combines the regression tree and the linear mixed effects model for modeling multivariate response longitudinal or clustered data. The Multivariate RE-EM tree method estimates a population-level single tree structure that is driven by the multiple responses simultaneously and object-level random effects for each response variable, where correlation between the response variables and between the associated random effects are each allowed. Through simulation studies, we verify the advantage of the Multivariate RE-EM tree over the use of multiple univariate RE-EM trees and the Multivariate Regression Tree. We apply the Multivariate RE-EM tree to analyze a real data set that contains multidimensional nonfinancial characteristics of poverty of different countries as responses, and various potential causes of poverty as predictors. 
\end{abstract}


\section{Introduction}

The regression tree is a nonparametric method for estimating a regression function. It has the advantages of being flexible, easily interpretable, well-suited to uncovering interactions among predictors, and having the ability to produce well-defined subgroups of the data via membership in the terminal nodes of the constructed tree. Tree-based methods have been developed primarily for situations with a univariate response variable \citep[see, e.g.,][]{breiman1984classification}, but they can be applied as well to multivariate response data. A multivariate regression tree provides two additional potential benefits to those for a univariate response: the ability to allow for correlation between the response variables, and the ability to impose the restriction of a single tree structure that is driven by the multiple responses simultaneously. Consider, for example, data with a bivariate blood pressure response (systolic and diastolic): a method using various predictors that determines a single set of subgroups of patients with ``similar blood pressure" based on both responses is likely to be much more useful than a method that produces two different sets of subgroups for the two responses. De'Ath \cite{de2002multivariate} proposed such an algorithm called \textit{multivariate regression trees} (MRT) that can construct a single tree for data with multiple response variables, implemented in R package $\mathtt{mvpart}$. MRT straightforwardly extends the univariate regression tree by defining a new impurity measure for multiple response variables. It can then split the tree nodes in the same way as the univariate regression tree based on this new measure.

The traditional tree-based methods, including MRT, assume independence over the observations. This assumption is violated in longitudinal or clustered data, as longitudinal structure often implies correlation among the observations within an object. Ignoring the structure can result in overstating the correlation in the population and negatively affecting model selection, often leading to models that are too complex. Therefore, the tree-based algorithms have to be specifically modified to tackle this issue. Segal \cite{segal1992tree} first extended the regression tree to a multiple response version that is designed for longitudinal data. However, the ``multiple response" in this method only refers to the outcomes of a single response variable at different time points. It considers a covariance structure among those outcomes instead of assuming them as independent, but does not allow for different response variables such as the blood pressure variables mentioned above. Hajjem et al. \cite{hajjemetal} proposed a \textit{mixed effects regression tree} (MERT) algorithm for clustered data, which divides the model into two parts: a fixed (population-averaged) effect and a random (cluster-specific) component. They use a standard regression tree, \textit{classification and regression tree} (CART), for the fixed effect and a linear structure for the random effects. Sela and Simonoff \cite{sela2012re} independently proposed a similar estimation method called RE-EM that also uses CART to estimate the fixed effects and a linear model for the random effects. They alternate between estimating the regression tree, assuming that estimates of the random effects are correct, and estimating the random effects, assuming that the regression tree for the fixed effects is correct. Both MERT and RE-EM are designed for univariate response data, which motivates us to extend their approach to a multivariate version by replacing CART with MRT. 

In this paper, we propose a regression tree method for longitudinal and clustered data with multiple responses. In Section \ref{sec:method}, we outline the algorithm for producing the tree and provide a generalized algorithm for non-Gaussian response variables. Section \ref{sec:MCsimul} provides discussion of Monte Carlo simulations investigating the performance of the method. In Section \ref{sec:poverty}, we examine a real data example that explores the multivariate nature of poverty, and Section \ref{sec:concl} concludes the paper.

\section{The Multivariate RE-EM Tree}
\label{sec:method}

To formally define the algorithm, we partially follow the notation in \cite{sela2012re}. Suppose there are $I$ objects $i=1,2,.., I$, and each object is observed at times $t=1,2,\dots,T_i$. Let $\bm{Y}_{it}=\left(Y^{(1)}_{it},\dots,Y^{(J)}_{it}\right)^{\top}$ denote the $J$ response variables of object $i$ at time $t$, and $\bm{X}_{it} = \left(X_{it1},\dots,X_{itk}\right)^{\top}$ denote the $k$ corresponding covariates. Given a design matrix $\bm{Z}_{it} \in \mathbb{R}^{q\times 1}$ at each observation, the multivariate model can be formalized as 
\begin{equation}
	\begin{aligned}
		\bm{Y}_{it} &= \bm{f}(\bm{X}_{it}) + \bm{B}_{i}\bm{Z}_{it} +  \bm{\varepsilon}_{it},
	\end{aligned}
	\label{model}
\end{equation}
where $\bm{f}(\bm{X})$ is a multivariate regression tree,  $\bm{B}_{i}$ is the random effect matrix of object $i$, and $\bm{\varepsilon}_{it}$ is the noise of observation $(i, t)$. Concretely, if we let $g_1,\dots,g_L$ denote the $L$ terminal nodes (i.e. leaves) of $\bm{f}$, and $\mu^{(j)}_{1},\dots,\mu^{(j)}_{L}$ denote the predicted response of $Y^{(j)}$ on each leaf, we can rewrite $\bm{f}(\bm{X})$ as  
\begin{equation}
	\bm{f}(\bm{X})= \left(\sum_{l=1}^L \mathbb{I} (\bm{X} \in g_l)\mu^{(1)}_{l},\, \dots \, ,\sum_{l=1}^L \mathbb{I} (\bm{X} \in g_l)\mu^{(J)}_{l}\right)^{\top},
	\label{model_tree}
\end{equation}
where $\mathbb{I}(\cdot)$ is the indicator function. $\bm{B}_{i}=\left(\bm{b}^{(1)}_{i},\dots,\bm{b}^{(J)}_{i}\right)^{\top} \in \mathbb{R}^{J\times q}$, where $\bm{b}^{(j)}_{i}$ stands for the time-constant, object-specific random effects of object $i$ on the $j$-th response variable. In particular, if we set $\bm{Z}_{it}=1$ and $\bm{B}_{i}=\left(b^{(1)}_{i},\dots,b^{(J)}_{i}\right)^{\top} \in \mathbb{R}^{J\times 1}$, only a random intercept for each object for each response variable will be included in the model. We assume \[\mathrm{vec}(\bm{B}_i)=\begin{pmatrix}\bm{b}^{(1)}_{i}\\ \vdots \\ \bm{b}^{(J)}_{i}\end{pmatrix} \sim \mathcal{N} \left(\bm{0},\bm{D}\right),\] 
which allows correlation between $\bm{b}^{(j)}_{i}$s. The noise $\bm{\varepsilon}_{it}=\left(\varepsilon^{(1)}_{it},\dots,\varepsilon^{(J)}_{it}\right)^{\top}$, and we suppose for each $i$,
\[ \begin{pmatrix}\bm{\varepsilon}_{i1}\\
	\bm{\varepsilon}_{i2}\\
	\vdots\\
	\bm{\varepsilon}_{iT_i}\end{pmatrix} \sim \mathcal{N} (\bm{0},\bm{R}_i).\]
In other words, we assume the noise random variables are independent across different objects, but within one object they are allowed to be correlated by setting non-diagonal $\bm{R}_i$.

We are now ready to define the multivariate RE-EM algorithm:

\begin{enumerate}
	\item[\textbf{Step 1}] Initialize $\widehat{\bm{B}}_i=\left(\widehat{\bm{b}}_i^{(1)}, \dots,\widehat{\bm{b}}_i^{(J)}\right)^{\top}=\bm{0}$ for each $i$.

	\item[\textbf{Step 2}] Iterate the following (a) and (b) until convergence (the increase of the likelihood function in step (b) being less than some tolerance
	value).
	
	(a) Fit an MRT with target variables $\left\lbrace\bm{Y}_{it}-\widehat{\bm{B}}_i\bm{Z}_{it}\right\rbrace$ 
	and predictors $\left\lbrace\bm{X}_{it}\right\rbrace$, $i\in\{1,2,...,I\}, t\in\{1,..., T_i\}$. Record the terminal nodes of this tree, denoted by $\widehat{g}_1,\dots,\widehat{g}_L$.
	
	(b) Fit a multivariate mixed linear model as follows,
	\[ \bm{Y}_{it}= \begin{pmatrix}
		\mu_1^{(1)}&\mu_2^{(1)}&
		\cdots&\mu_L^{(1)}\\
		\vdots & \vdots  & \ddots   & \vdots \\
		\mu_1^{(J)}&\mu_2^{(J)}&\cdots&\mu_L^{(J)}
	\end{pmatrix}\begin{pmatrix}
		\mathbb{I} (\bm{X}_{it} \in \widehat{g}_1)\\
		\vdots\\ 
		\mathbb{I} (\bm{X}_{it} \in \widehat{g}_L)
	\end{pmatrix} + \bm{B}_i\bm{Z}_{it}  + 
	\bm{\varepsilon}_{it}.\]
	Extract the estimated random effects $\widehat{\bm{B}}_{i}$ from this model.
	
	\item[\textbf{Step 3}] Take the mixed linear model in the last iteration to be the final result, with parameters $\widehat{g}_l, \widehat{\mu}^{(j)}_{l}, \widehat{\bm{B}}_{i}$.
	
\end{enumerate}

The MRT algorithm proposed in \cite{de2002multivariate} uses Euclidean distance for measuring the diversity within a tree node. Since the response variables may have different scales, we should potentially standardize them at the beginning of the algorithm. For pruning, the multivariate RE-EM tree uses $k$-fold cross-validation in each iteration to select the tree, using the selecting method ``minimum" or ``one-standard-error." More details about standardization and pruning will be given in Section \ref{sec:methods}. 

We further extend the multivariate RE-EM tree algorithm to handle datasets with non-Gaussian response variables, using the \textit{penalized quasi-likelihood} (PQL) method investigated by \cite{breslow1993approximate} for \textit{generalized linear mixed models} (GLMM), as was done for univariate response trees in \cite{hajjem2017generalized}. As a generalization of \eqref{model}, assume that for all $i,t,j$,
\[\mathbb{E}\left(Y^{(j)}_{it}\mid \bm{b}^{(j)}_{i}\right) =  h\left(\eta_{it}^{(j)}\right), \quad \eta_{it}^{(j)}:=f^{(j)}(\bm{X}_{it}) + \bm{Z}_{it}^{\top}\bm{b}_i^{(j)}, \]
where $h$ is a known function and $f^{(j)}$ denotes the $j$-th response of a fixed multivariate tree $\bm{f}$. Given an estimator $\widehat{\eta}_{it}^{(j)}$, the first-order Taylor Series expansion $\mathbb{E}\left(Y^{(j)}_{it}\mid \bm{b}^{(j)}_{i}\right)\approx h\left(\widehat{\eta}^{(j)}_{it}\right) + h'\left(\widehat{\eta}^{(j)}_{it}\right)\left(\eta^{(j)}_{it}-\widehat{\eta}^{(j)}_{it}\right)$ implies that
\begin{equation}
	\label{eq:pseudo-response}
	\widetilde{Y}_{it}^{(j)}:=\widehat{\eta}^{(j)}_{it}+\left[h'\left(\widehat{\eta}^{(j)}_{it}\right)\right]^{-1}\left(Y_{it}^{(j)}-h\left(\widehat{\eta}^{(j)}_{it}\right)\right)\approx \eta_{it}^{(j)} =f^{(j)}(\bm{X}_{it}) + \bm{Z}_{it}^{\top}\bm{b}_i^{(j)}.
\end{equation}
Therefore, the RE-EM algorithm can be applied to the pseudo-response variables $\widetilde{Y}_{it}^{(j)}$.
The generalized multivariate RE-EM tree algorithm is formally presented as follows.

\begin{enumerate}
	\item[\textbf{Step 1}] Initialize $\widehat{\bm{B}}_i=\left(\widehat{\bm{b}}_i^{(1)}, \dots,\widehat{\bm{b}}_i^{(J)}\right)^{\top}$ for each $i$ and $\widehat{\eta}_{it}^{(j)}$ for each $i,t,j$. Compute the pseudo-response variables $\widetilde{\bm{Y}}_{it}=\left(\widetilde{Y}_{it}^{(1)},\dots, \widetilde{Y}_{it}^{(J)}\right)^{\top}$ by \eqref{eq:pseudo-response}.

	\item[\textbf{Step 2}] Iterate the following (a), (b) and (c) until convergence.
	
	(a) Fit an MRT with target variables $\left\lbrace\widetilde{\bm{Y}}_{it}-\widehat{\bm{B}}_i\bm{Z}_{it}\right\rbrace$ 
	and predictors $\left\lbrace\bm{X}_{it}\right\rbrace$, $i\in\{1,2,...,I\}, t\in\{1,..., T_i\}$. Record the terminal nodes of this tree, denoted by $\widehat{g}_1,\dots,\widehat{g}_L$.
	
	(b) Fit a multivariate mixed linear model as follows,
	\[ \widetilde{\bm{Y}}_{it}= \begin{pmatrix}
		\mu_1^{(1)}&\mu_2^{(1)}&
		\cdots&\mu_L^{(1)}\\
		\vdots & \vdots  & \ddots   & \vdots \\
		\mu_1^{(J)}&\mu_2^{(J)}&\cdots&\mu_L^{(J)}
	\end{pmatrix}\begin{pmatrix}
		\mathbb{I} (\bm{X}_{it} \in \widehat{g}_1)\\
		\vdots\\ 
		\mathbb{I} (\bm{X}_{it} \in \widehat{g}_L)
	\end{pmatrix} + \bm{B}_i\bm{Z}_{it}  + 
	\bm{\varepsilon}_{it}.\]
	
	(c) Update $\widehat{\eta}_{it}^{(j)}= \sum_{l=1}^{L} \widehat{\mu}^{(j)}_{l} \mathbb{I} (\bm{X}_{it} \in \widehat{g}_l)+ \bm{Z}_{it}^{\top}\widehat{\bm{b}}^{(j)}_i$ and $\widetilde{\bm{Y}}_{it}$ correspondingly.
	
	\item[\textbf{Step 3}] Take the mixed linear model in the last iteration to be the final result, with parameters $\widehat{g}_l, \widehat{\mu}^{(j)}_{l}, \widehat{\bm{B}}_{i}$.
	
\end{enumerate}

Note that in addition to providing a regression tree for categorical response variables with non-Gaussian distributions such as the binomial and Poisson, this algorithm also provides a version of a classification tree for two-category (binary) response variables, as they can be fit assuming a Bernoulli distribution.

\section{Simulation Results}
\label{sec:MCsimul}

\subsection{Bivariate RE-EM with simple tree structure}\label{sec:bi-simple}

We first set $\bm{f}(\bm{X})$ to be a relatively simple structure and $J=2$ to compare the multivariate RE-EM tree to other competitor methods listed in Section \ref{sec:methods}. 

\subsubsection{Settings} \label{sec:setting}

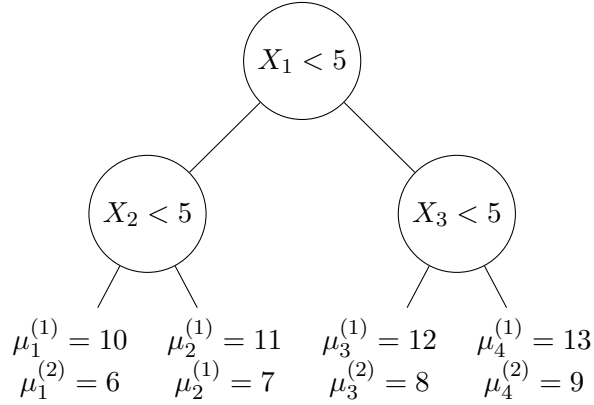
\begin{figure}[!t] 
	\centering   
	\begin{forest} 
		[$X_1<5$, circle,draw
		[$X_2<5$, circle, draw
		[{$\mu^{(1)}_1=10$}\\ {$\mu^{(2)}_1=6$}, align=center, base=bottom]
		[{$\mu^{(1)}_2=11$}\\ {$\mu^{(1)}_2=7$}, align=center, base=bottom]
		]
		[$X_3<5$, circle, draw
		[{$\mu^{(1)}_3=12$}\\ {$\mu^{(2)}_3=8$}, align=center, base=bottom]
		[{$\mu^{(1)}_4=13$}\\ {$\mu^{(2)}_4=9$}, align=center, base=bottom]
		]
		]
	\end{forest}
	\caption{Simple tree structure, as discussed in section \ref{sec:setting}}\label{fig:simpletree}
\end{figure}

We generate simulated data sets with $I=50, 100, 200, 400, 800$ and $T_i=5, 10, 25, 50$ for all $i$. Following the procedure in \cite{fu2015unbiased}, we first generate $X_1,X_2,X_3,X_4,X_5,X_6 \sim Unif(0,10)$ (i.i.d.), then  round $X_6$ to the nearest integer. Additionally, we generate another predictor $X_7 \sim Unif\{0,5.77\}$, which has the same variance as $X_1,\dots,X_5$. The tree structure is set as follows (using the notation defined in \eqref{model}, also shown in Figure \ref{fig:simpletree}): 
\begin{align*}
	g_1&=(X_1 \leq 5) \wedge (X_2 \leq 5), \mu^{(1)}_{1} = 10, \mu^{(2)}_{1} = 6.\\
	g_2&=(X_1 \leq 5) \wedge (X_2 > 5), \mu^{(1)}_{2} = 11, \mu^{(2)}_{2} =7.\\
	g_3&=(X_1 > 5) \wedge (X_3 \leq 5), \mu^{(1)}_{3} = 12, \mu^{(2)}_{3} = 8.\\
	g_4&=(X_1 > 5) \wedge (X_3 > 5), \mu^{(1)}_{4} = 13, \mu^{(2)}_{4} = 9.
\end{align*}

For the random effect, we take the design matrix $\bm Z_{it}=1$, with
$\begin{pmatrix}\bm{b}^{(1)}_{i}\\ \bm{b}^{(2)}_{i}\end{pmatrix} \sim \mathcal{N} \left(\bm{0},\bm{D}\right)$ and $\bm{D}=\begin{pmatrix}
	1& \sigma_{12}\\\sigma_{12} & 1
\end{pmatrix}$, where  $\sigma_{12}=0$, $.25$, $.5$, $.75$. The covariance matrix of noise $\bm{R}_i$ is set to be $ \begin{pmatrix}
	\sigma^2_\varepsilon& 0\\0 & \sigma^2_\varepsilon
\end{pmatrix}$ with $\sigma^2_\varepsilon=.5$, $1$, $1.5$, $2$. Based on this setting, the response values of $\bm{Y}$ are generated by \eqref{model} and \eqref{model_tree}. 

For each $(I, \sigma_{12}, \sigma_\varepsilon)$, we generate a test set containing the same $I$ objects as in the training set, each with 20 new observations. Concretely, we keep $\bm{b}^{(1)}_{i}, \bm{b}^{(2)}_{i}$ to be the same as the training set for all $i$, then randomly sample $\bm{X}_{it_{\mathrm{test}}}$ and $\bm{\varepsilon}_{it_{\mathrm{test}}}$ for $t_{\mathrm{test}}=1,2,...,20$ from the same distribution as above.

\subsubsection{Methods for Comparison}\label{sec:methods}

For comparison, we apply the following 4 competitive methods on the simulation sets:

\begin{enumerate}
	\item[1.] Separate univariate linear mixed models, referred to as ``unilme." This method only fits a linear mixed model of each response on the covariates, and does not account for correlation within $\bm{B}_i$.
	
	\item[2.] One multivariate linear mixed model, referred to as ``multilme." This method still imposes linearity, but allows correlation between the responses.
	
	\item[3.] One multivariate tree (MRT) that doesn't account for the longitudinal structure, referred to as ``multitree."

	\item[4.] Separate univariate RE-EM trees, referred to as ``uniREEM." This is similar to method 1, but removes the linear model restriction. Note that this doesn't impose a restriction that there is a single tree structure driving the response variables simultaneously, which potentially adds noise if there actually is a single tree structure driving both responses.
\end{enumerate}

For the multivariate RE-EM tree algorithm, we also compare four versions, with two standardization methods and two criteria for selecting the tree after the $k$-fold cross-validation (we choose $k=10$ as the default value). The standardization methods are 

\begin{enumerate}
	\item ``marg": Marginalizing each response variable $Y^{(j)}$ by 
	\[Y_{it, \mathrm{scaled}}^{(j)}\leftarrow \frac{Y_{it}^{(j)}-\mathrm{mean}_{i,t}\left(Y_{it}^{(j)}\right)}{\mathrm{sd}_{i,t}\left(Y_{it}^{(j)}\right)}.\]
	\item ``cov": Estimate the covariance matrix $\mathrm{cov}(\bm Y)$ at the beginning, where $\bm Y=\left(Y^{(1)},...,Y^{(J)}\right)$. Let
	\[\bm Y_{\mathrm{scaled}} \leftarrow \bm Y \left(\mathrm{cov}(\bm Y)\right)^{-1/2}.\]
	In this way, the Euclidean distance computed by the Multivariate Regression Tree will be actually the Mahalanobis distance.
\end{enumerate}

The choosing criteria are

\begin{enumerate}
	
	\item ``min": Choosing the tree with minimum estimated predictive error based on $k$-fold
	cross-validation. 
	
	\item ``1se": Choosing the simplest tree with cross-validated estimated predictive error within one standard error of the minimized estimated predictive error.
	
\end{enumerate}

The main measure to evaluate the performance of the models is the object-level Predicted Mean Squared Error (PMSE), which is defined by
\[ \widehat{PMSE} = \frac{1}{20I} \sum_{i, t_{\mathrm{test}},j}\left(\widehat{Y}^{(j)}_{it_{\mathrm{test}}}-Y^{(j)}_{it_{\mathrm{test}}}\right)^2.\]
We also report the EMSE (Estimation Mean Squared Error) of the population-level fixed effect for all the methods. Specifically, for all methods except ``multitree", given a fitted linear mixed model $\bm{Y}_{it}=\widehat{\bm{f}}(\bm{X}_{it})+\widehat{\bm{B}}_i\bm{Z}_{it}$ and a new data point $(\bm{X}_0,\bm{Z}_{i_0t_0})$ $(i_0 \in \{1,2,..,I\})$, the estimated fixed effect at the new point is $\widehat{\bm{f}}(\bm{X}_{0})$ and the predicted value $\widehat{\bm Y}_{it_{\mathrm{test}}} = \widehat{\bm{f}}(\bm{X}_0) + \widehat{\bm{B}}_{i_0}\bm{Z}_{i_0t_0}$. For ``multitree", the estimated fixed effect and the predicted value are the same since there is no random effect. Further, we also record the tree-recovering rates of the tree-based methods, i.e., the rate that a tree method can capture the tree structure (the shape of the tree and the variable on which each split is based) correctly. In the ``uni-REEM" method, we record the recovering rates of the $J$ univariate RE-EM trees separately.  The final results are averaged over 400 repeats.


\subsubsection{Results for Simpler Bivariate Tree Structure}

\begin{figure}[!htp]
	\centerline{
		\includegraphics[width=1.0\linewidth, height=0.45\textheight]{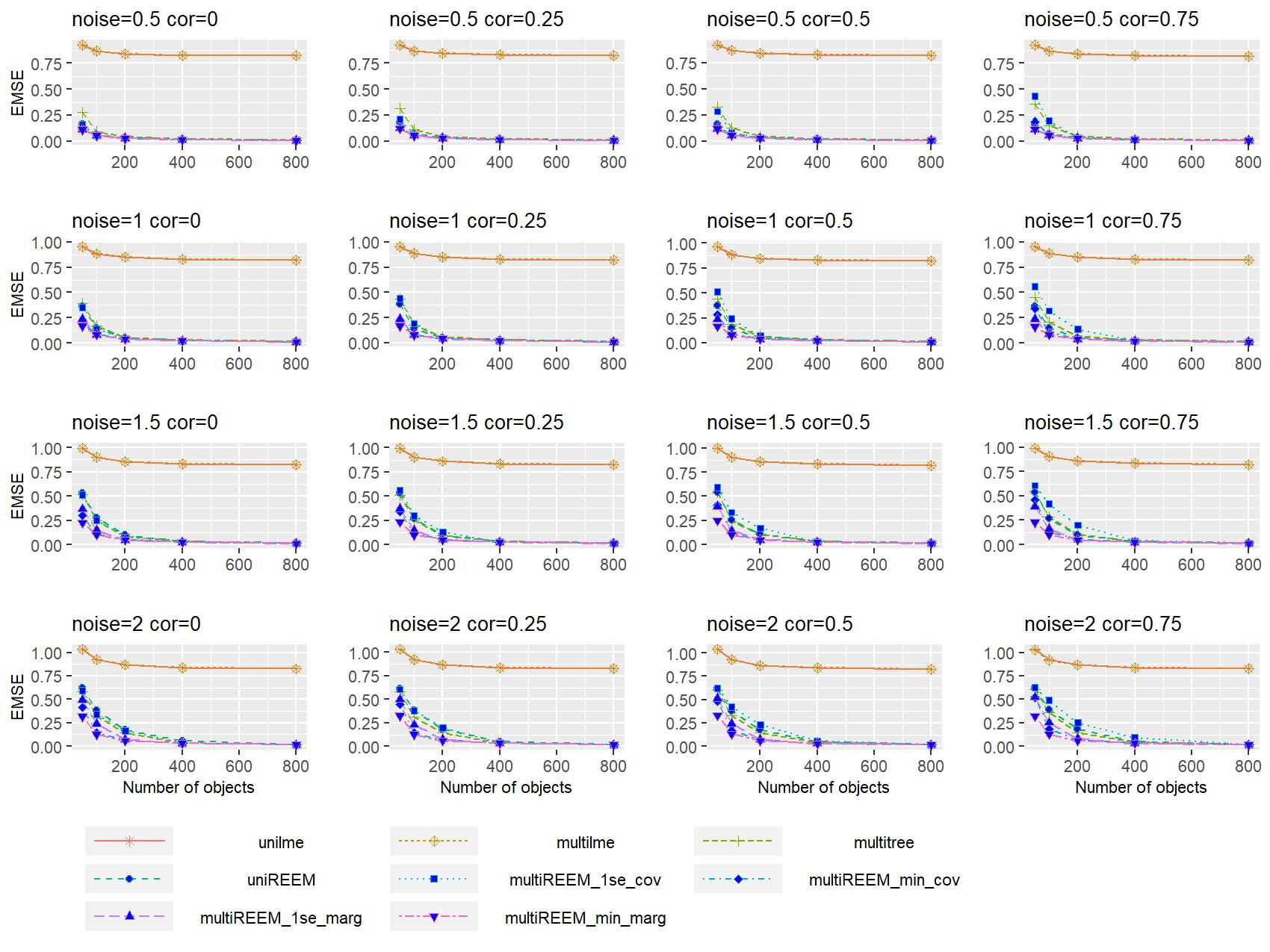}}
	\caption{EMSE in population for $T_i=5$ for the simpler bivariate tree. Note that the lines for separate univariate linear mixed effects models (``unilme") and a single multivariate linear mixed effects model (``multilme") are almost the same.}
	\label{fig:ti5-str1-fixed}
\end{figure}

\begin{figure}[!htp]
	\centerline{
		\includegraphics[width=1.0\linewidth, height=0.45\textheight]{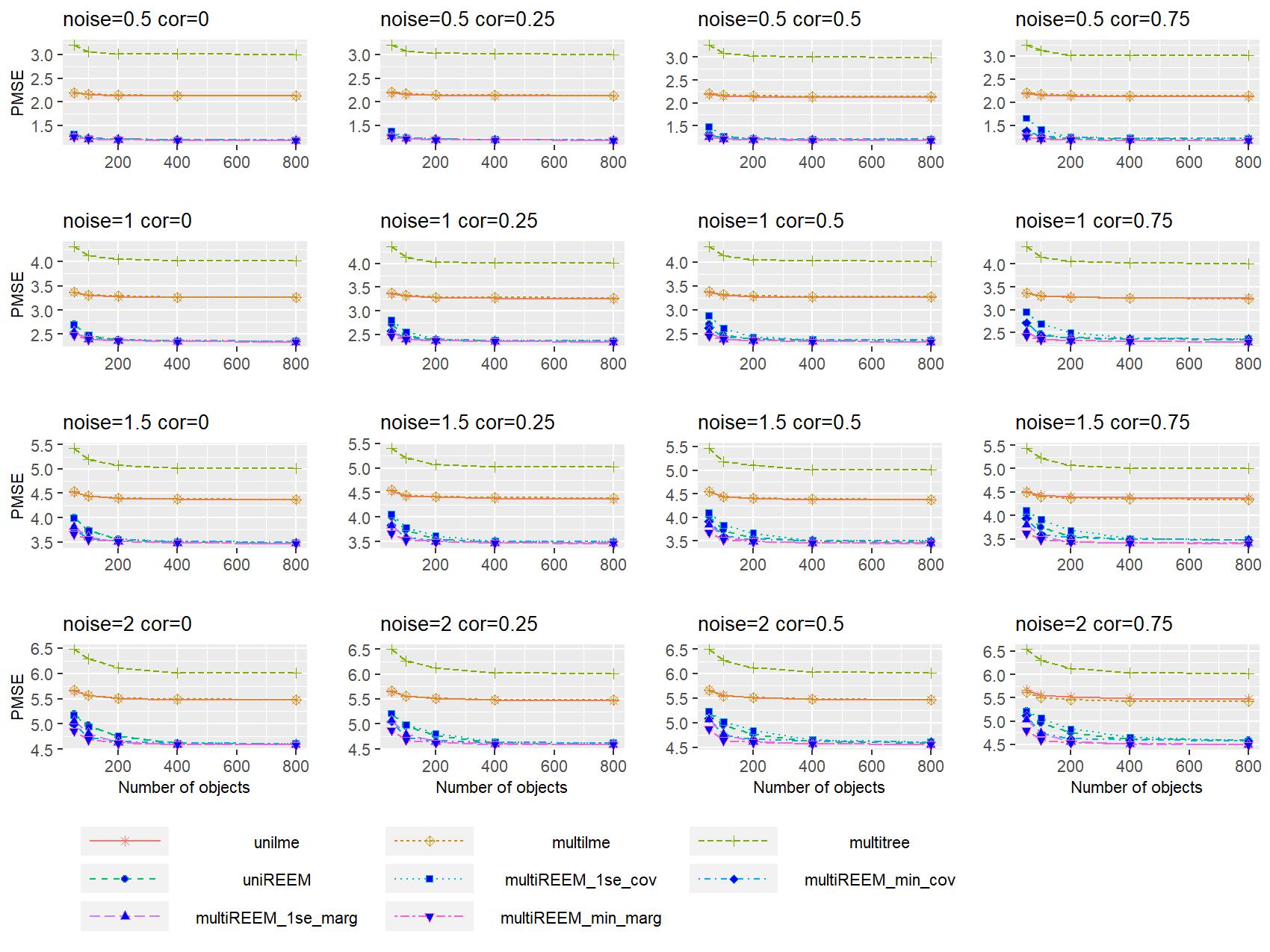}}
	\caption{PMSE in object of all the methods for $T_i=5$ with the simpler bivariate tree structure}
	\label{fig:ti5-str1-group-all}
\end{figure}

The results for the estimation in population and prediction in object for all of the methods when $T_i=5$ are shown in Figures \ref{fig:ti5-str1-fixed} and \ref{fig:ti5-str1-group-all}, respectively. In both figures, it is obvious that the tree-based methods performed better than the linear methods. Figure \ref{fig:ti5-str1-fixed} indicates that all of the tree methods are competitive for population-level estimation. This illustrates that the dominant factor in estimating the population-level expected response is to recognize it as a tree; beyond that, doing so with a tree that accounts for the longitudinal structure provides an advantage over one that does not account for it, but this effect is virtually gone when the number of objects $I$ exceeds 200 to 400. For this reason, we will focus on object-level prediction here. 

For the object-level prediction in Figure \ref{fig:ti5-str1-group-all}, both of the versions of the proposed algorithm ``multiREEM" with ``marg" standardization outperform the other methods, including ``multitree" and ``uniREEM'', especially when the correlation $\sigma_{12}$ or the noise $\sigma_\varepsilon$ is large. As would be expected, the trees that account for the random effects clearly outperform the tree that ignores the longitudinal structure when attempting prediction at the level of the individual, so the latter will not be considered further.

\begin{figure}[!htp]
	\centerline{
		\includegraphics[width=1.0\linewidth, height=0.45\textheight]{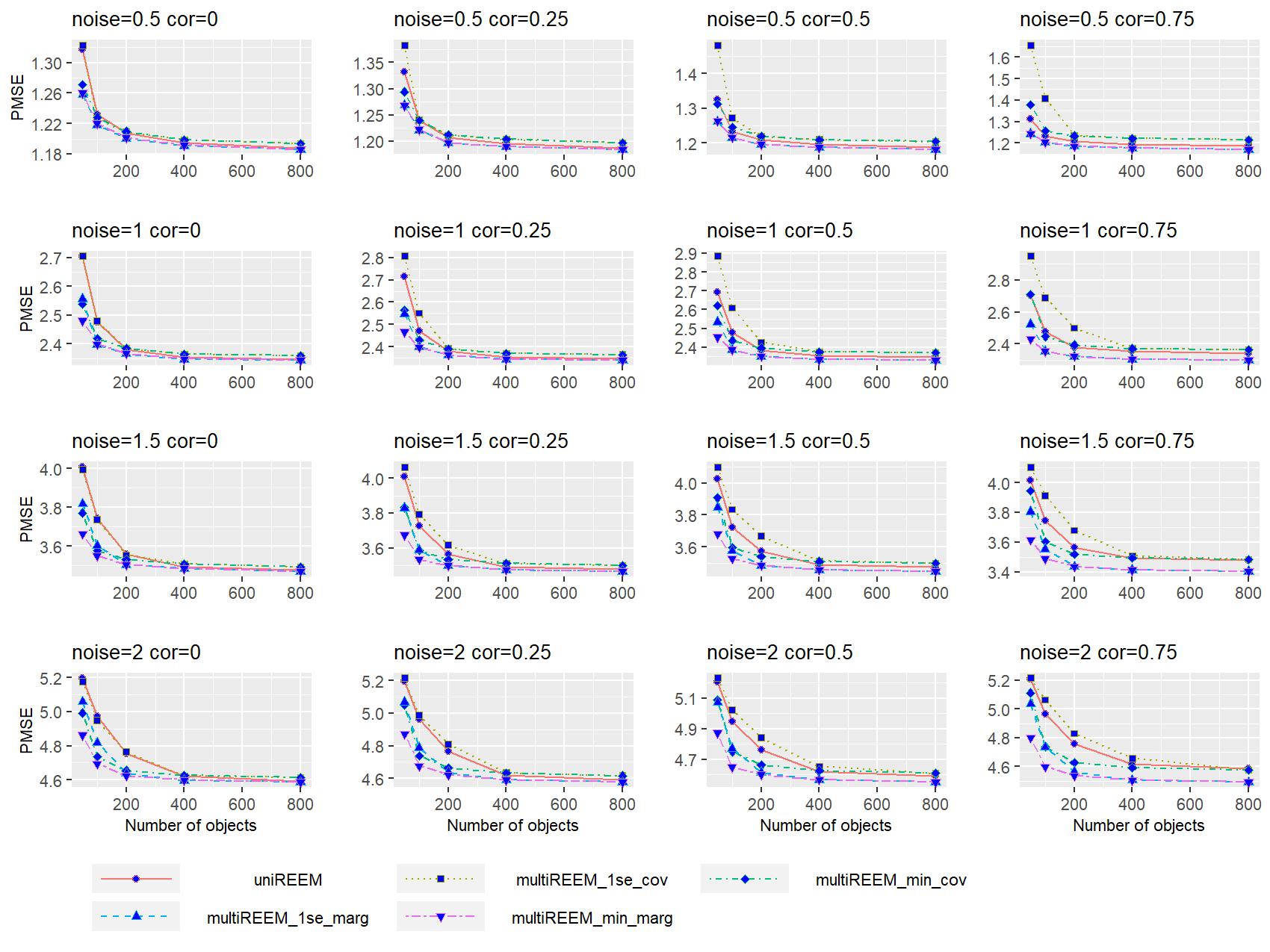}}
	\caption{PMSE in object of the RE-EM methods for $T_i=5$ with the simpler bivariate tree structure}
	\label{fig:ti5-str1-group-part1}
\end{figure}

\begin{figure}[!htp]
	\centerline{
		\includegraphics[width=1.0\linewidth, height=0.45\textheight]{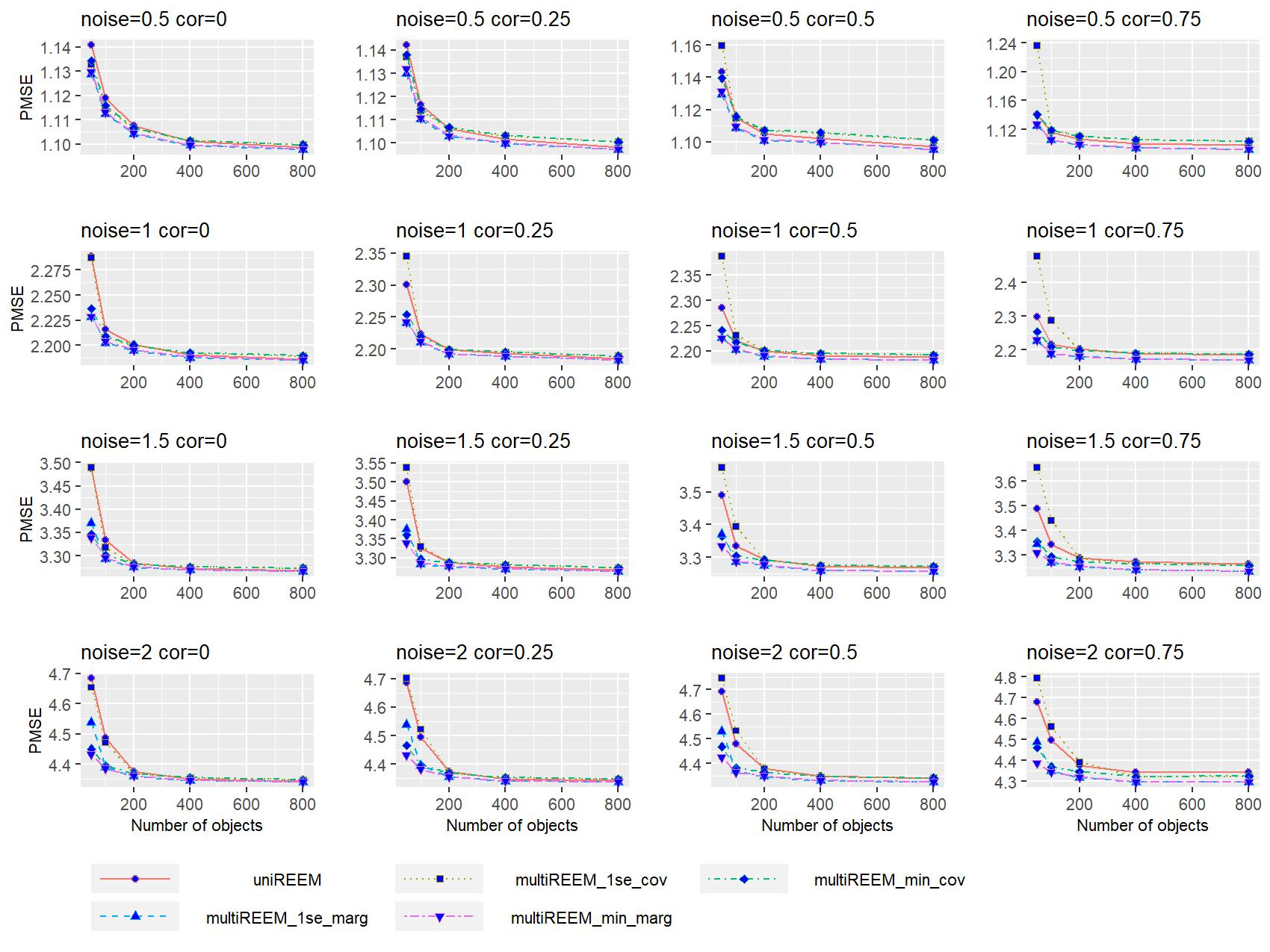}}
	\caption{PMSE in object for $T_i=10$ with the simpler bivariate tree structure}
	\label{fig:ti10-str1-group-part1}
\end{figure}

\begin{figure}[!htp]
	\centerline{
		\includegraphics[width=1.0\linewidth, height=0.45\textheight]{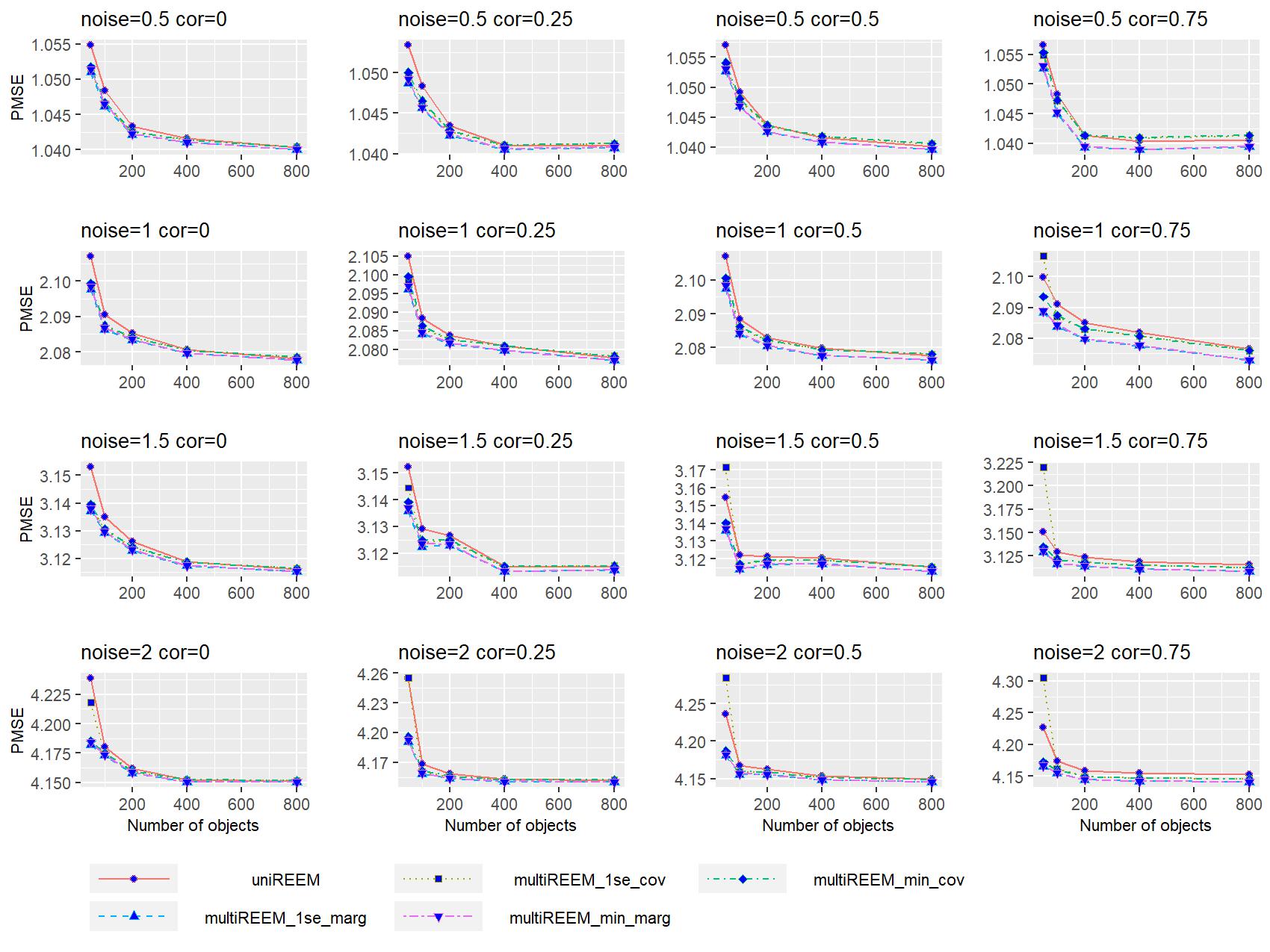}}
	\caption{PMSE in object for $T_i=25$ with the simpler bivariate tree structure}
	\label{fig:ti25-str1-group-part1}
\end{figure}

\begin{figure}[!htp]
	\centerline{
		\includegraphics[width=1.0\linewidth, height=0.45\textheight]{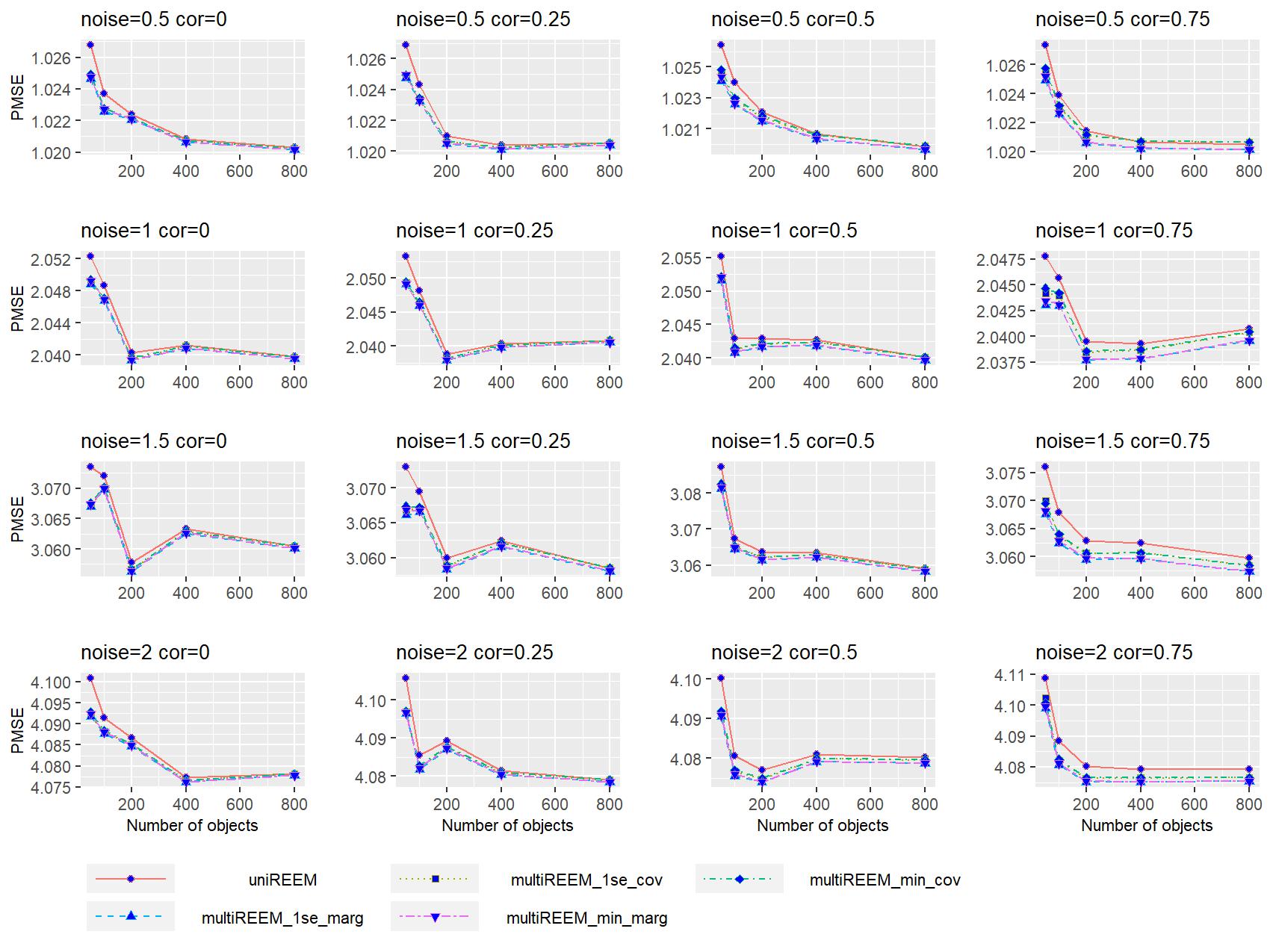}}
	\caption{PMSE in object for $T_i=50$ with the simpler bivariate tree structure}
	\label{fig:ti50-str1-group-part1}
\end{figure}

To see the differences among the tree-based methods more clearly, we discarded linear methods ``unilme" and ``multilme" in the following figures. The object-level PMSE for $T_i=5, 10, 25, 50$ are shown in Figures \ref{fig:ti5-str1-group-part1}--\ref{fig:ti50-str1-group-part1}. In almost all of the cases, the multiREEM tree ``min\_marg" and ``1se\_marg" perform the best, with no meaningful difference between them. The other standardizing method ``cov" does not perform very well, and ``1se\_cov" is always worse than ``min\_cov". For $T_i= 5, 10, 25$, the multiREEMtree ``1se\_cov" is never better than the ``uniREEM" either for small $I$, small $\sigma_{12}$ or small $\sigma_\varepsilon$. Only when $I, \sigma_{12}, \sigma_\varepsilon$ are all large enough, or $T_i=50$, can the ``cov" standardization show an advantage over ``uniREEM", but it is still worse than the ``marg" standardization. 

The better performance of the ``min\_cov" pruning versus the ``1se\_cov" pruning suggests that the covariance standardization tends to result in fitted trees that are too simple. This is further supported by simulations based on a more complex fixed effects structure, in which the ``cov" standardization performs even more poorly; for this reason, it will not be considered further in the results presented here.

Using the ``min" criterion, the advantage of fitting a multivariate tree versus two separate univariate trees when the true underlying structure is a single tree is evident from the figures, as the univariate RE-EM trees always lag behind in performance, sometime by a considerable amount. Clearly, avoiding the additional noise of fitting two separate longitudinal trees provides the potential of meaningful gains in predictive accuracy.

As would be expected, predictive performance improves with increasing number of objects $I$, and increasing number of time points $T_i$, reflecting both better estimation of the fixed effects and better prediction of the random effects.

The results are fairly insensitive to the presence of correlation in the random effects. Separate univariate RE-EM trees do not use that information, of course, and their performance suffers accordingly. Interestingly, the ``cov" standardization is also negatively affected by correlation, presumably because correlation of random effects is not the same as correlation between responses.

\begin{figure}[!htp]
	\centerline{
		\includegraphics[width=1.0\linewidth, height=0.45\textheight]{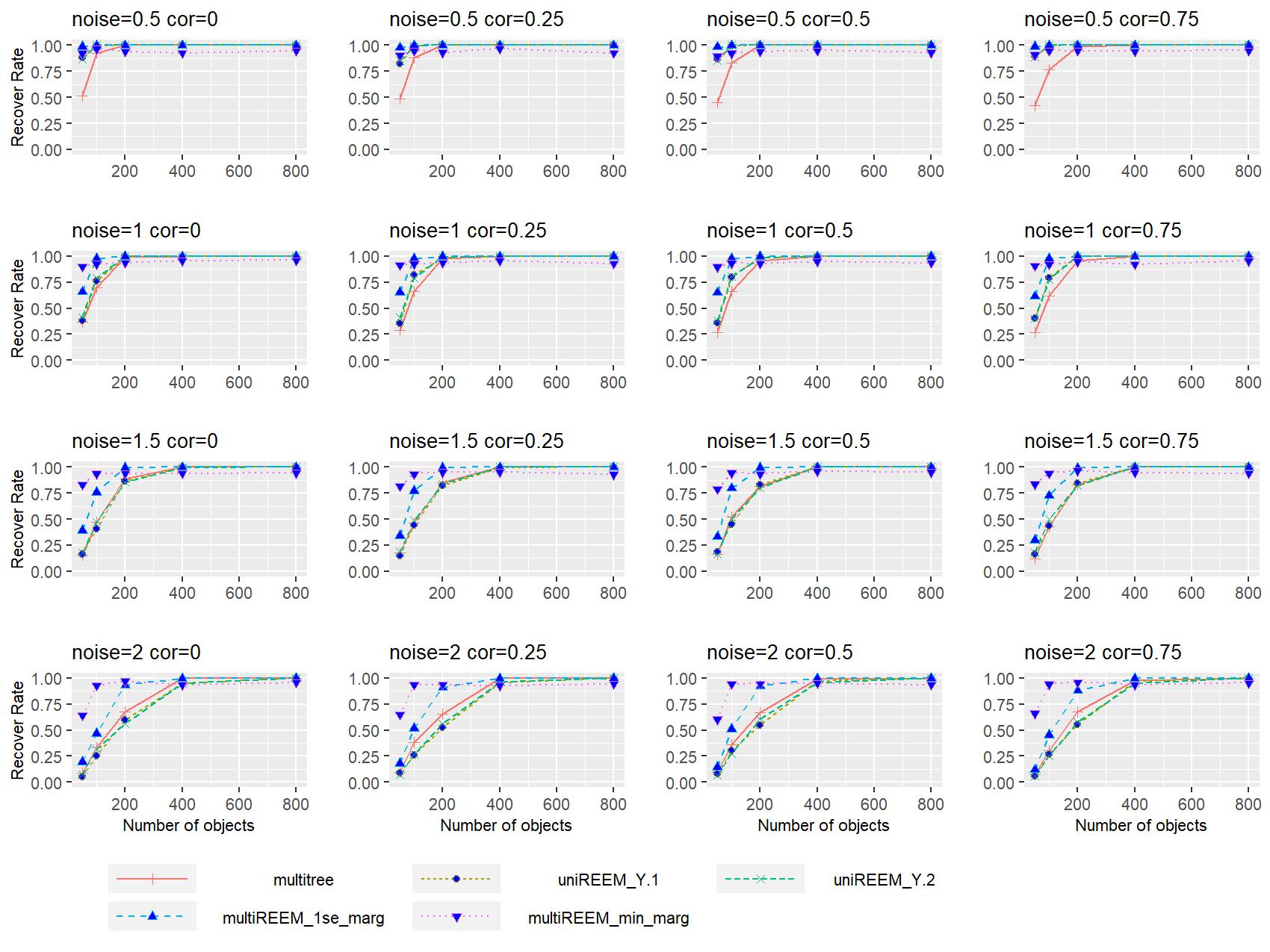}}
	\caption{Recovering rate for $T_i=5$ for the simpler bivariate tree structure}
	\label{fig:ti5-str1-rec}
\end{figure}
\begin{figure}[!htp]
	\centerline{
		\includegraphics[width=1.0\linewidth, height=0.45\textheight]{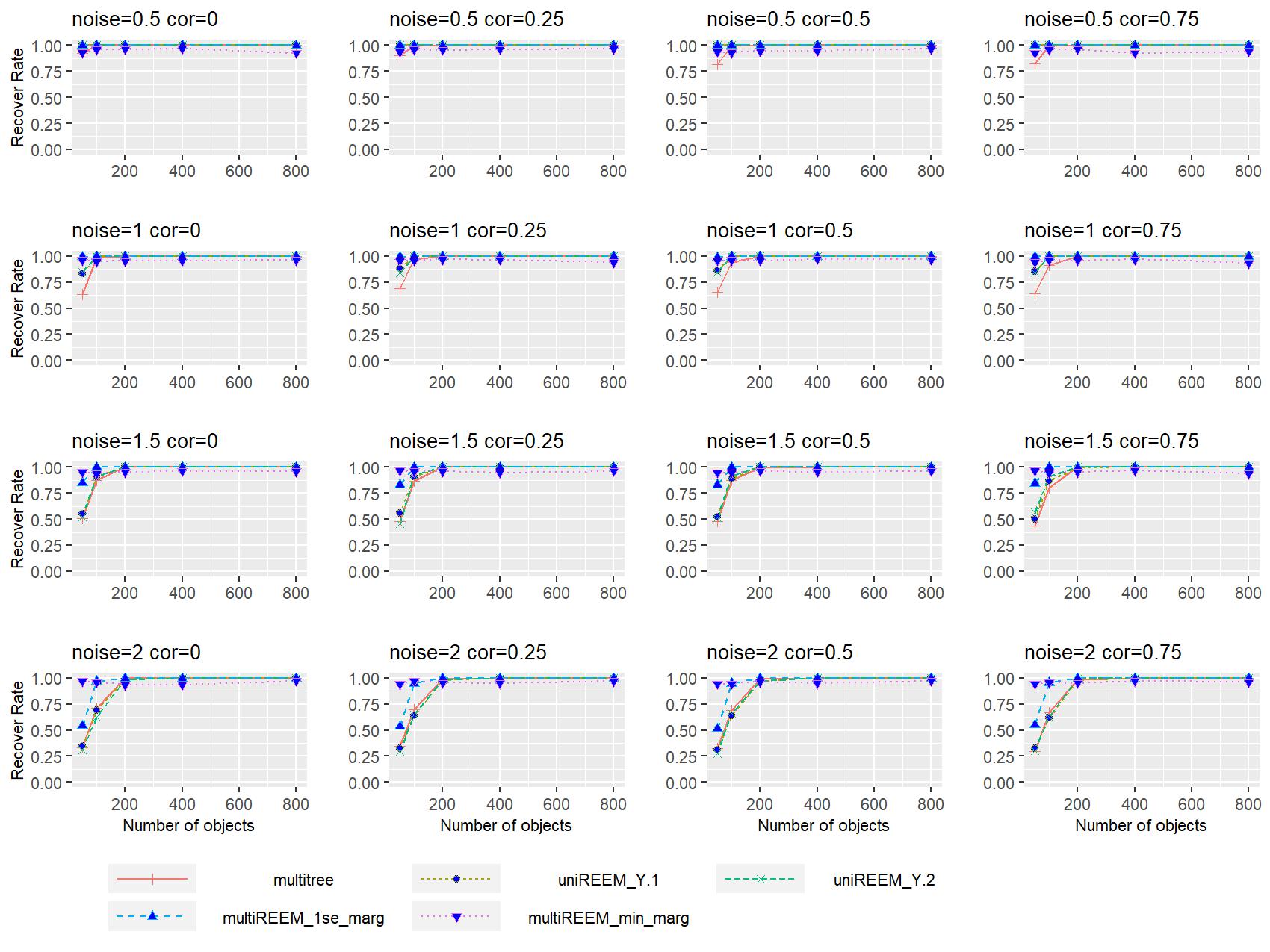}}
	\caption{Recovering rate for $T_i=10$ for the simpler bivariate tree structure}
	\label{fig:ti10-str1-rec}
\end{figure}
\begin{figure}[!htp]
	\centerline{
		\includegraphics[width=1.0\linewidth, height=0.45\textheight]{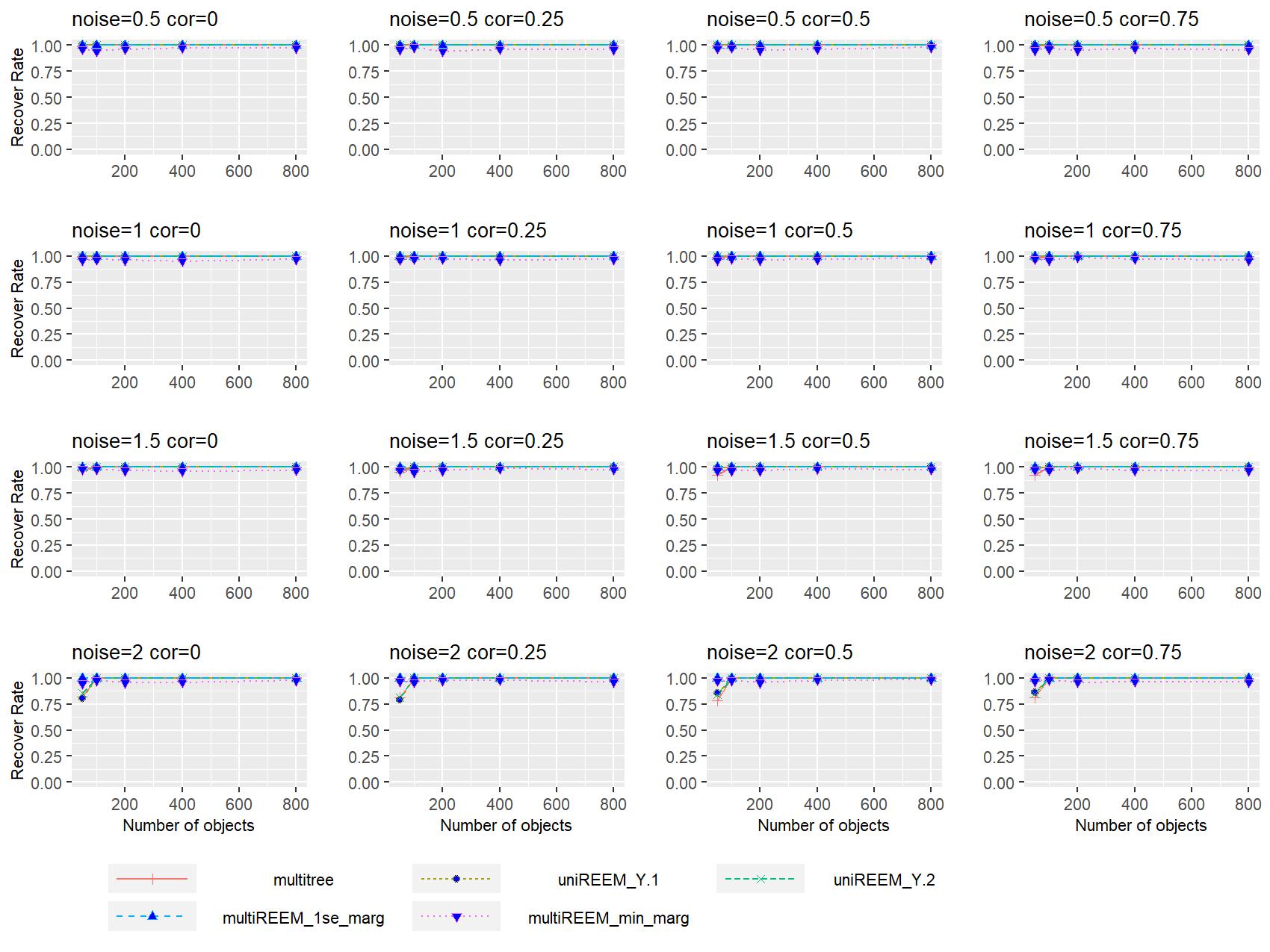}}
	\caption{Recovering rate for $T_i=25$ for the simpler bivariate tree structure}
	\label{fig:ti25-str1-rec}
\end{figure}
\begin{figure}[!htp]
	\centerline{
		\includegraphics[width=1.0\linewidth, height=0.45\textheight]{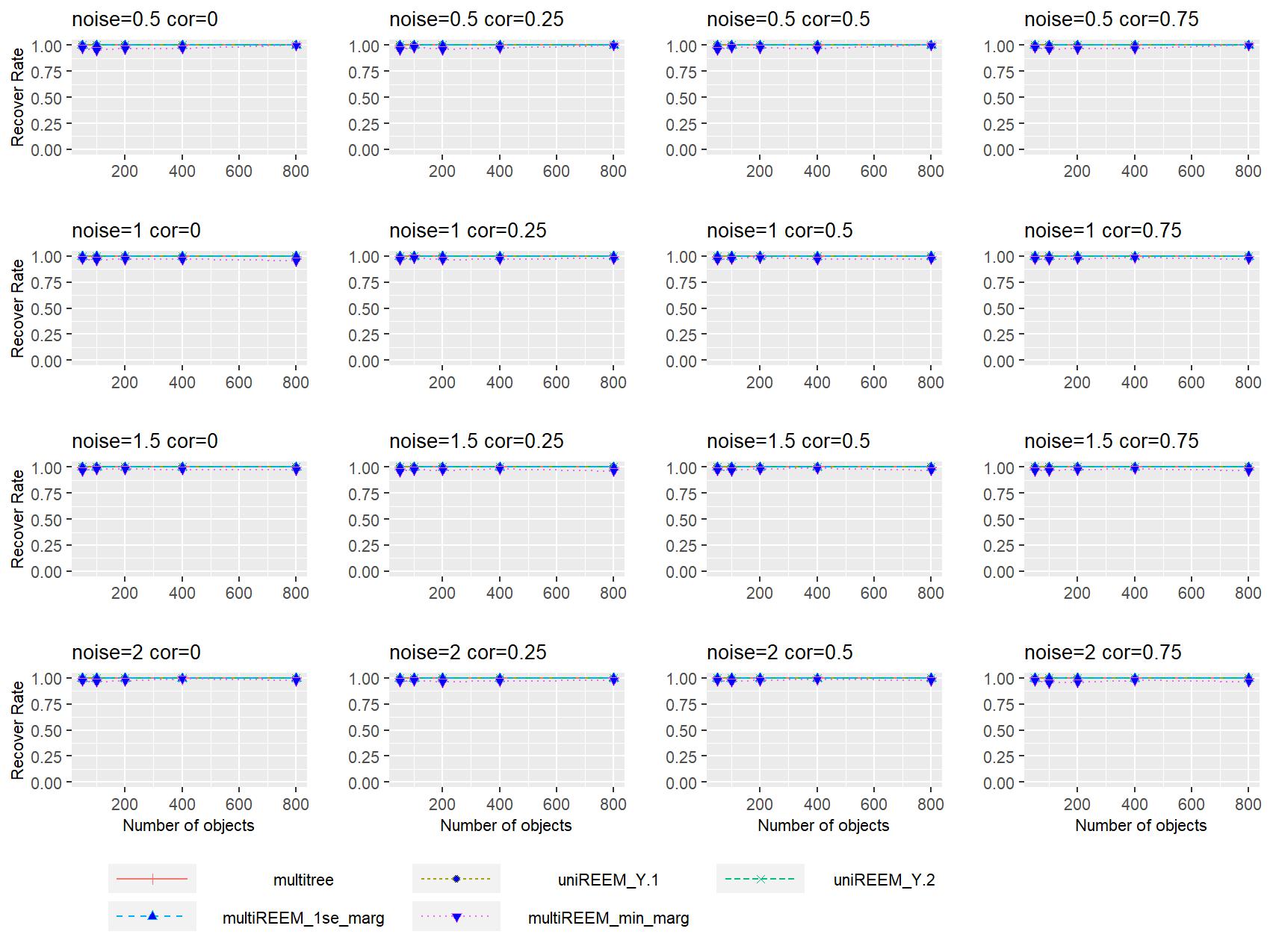}}
	\caption{Recovering rate for $T_i=50$ for the simpler bivariate tree structure}
	\label{fig:ti50-str1-rec}
\end{figure}

The tree recovering rates for the Multivariate Regression Tree, Univariate RE-EM trees and Multivariate RE-EM tree with marginal normalization are shown in Figures \ref{fig:ti5-str1-rec}--\ref{fig:ti50-str1-rec}. When $I$ is large enough, the multivariate RE-EM tree methods with marginal normalization and ``1se" criterion always have the highest recovering rate, while the ``min" criterion does not achieve this even when $I=800$. This implies that choosing the tree based on the minimum CV error may lead to overfitting, although this does not impact the Prediction Error, and the ``one-standard-error" choosing strategy overcomes this difficulty. However, when the noise is large and the sample size $I$ is small, the ``min" criterion achieves the best recovering rate compared to other competitors. It is striking that the univariate trees achieve perfect recovery of the underlying tree structure for a large enough number of objects and/or time points, but still typically lag behind the multivariate trees in terms of PMSE. This is because of both poorer estimation of fixed effects and poorer prediction of random effects; see Figures 5--8 in Section A.2 of the Appendix. 

Section A.2 of the Appendix summarizes the performances of the estimate of the correlation of the random effects $\sigma_{12}$ for different methods. As would be expected, accuracy improves as the number of objects increases. The estimation errors of the Multivariate RE-EM tree with ``marginal" standardization are always significantly lower than when using the multivariate linear mixed effects model, and the Multivariate RE-EM tree with ``covariance" standardization performs better when the true correlation is larger.

Section A.3 of the Appendix compares the performances of the different tree methods when there are no random effects (i.e., ${\bm B}=0$), and therefore the ordinary MRT is appropriate. The MRT and the multivariate RE-EM tree with ``marginal" standardization always outperform other methods, with no meaningful difference between them. This verifies that the proposed algorithm works as well as MRT (its special case with ${\bm B}=0$) in no-random-effect cases.

\subsection{A More Complex Tree Structure}\label{sec:bi-complex}

\begin{figure}[!t] 
	\centering   
	\begin{forest} 
		[$X_1<5$, circle,draw
		[$X_2<5$, circle, draw
		[$X_4<5$, circle, draw
		[{$\mu^{(1)}_1=6$}\\ {$\mu^{(2)}_1=4.5$}, align=center, base=bottom]
		[{$\mu^{(1)}_2=8$}\\ {$\mu^{(1)}_2=6.5$}, align=center, base=bottom]
		]
		[$X_5<5$, circle, draw
		[{$\mu^{(1)}_3=10$}\\ {$\mu^{(2)}_3=8.5$}, align=center, base=bottom]
		[{$\mu^{(1)}_4=12$}\\ {$\mu^{(1)}_4=10.5$}, align=center, base=bottom]
		]
		]
		[$X_3<5$, circle, draw
		[{$\mu^{(1)}_5=14$}\\ {$\mu^{(2)}_5=10.5$}, align=center, base=bottom]
		[$X_6<5$, circle, draw
		[{$\mu^{(1)}_6=16$}\\ {$\mu^{(2)}_6=12.5$}, align=center, base=bottom]
		[{$\mu^{(1)}_7=18$}\\ {$\mu^{(2)}_7=14.5$}, align=center, base=bottom]
		]
		]
		]
	\end{forest}
	\caption{A deeper and asymmetrical tree structure for section \ref{sec:bi-complex}. The true value of the $j$-th response variable on the $l$-th terminal node are given by $\mu_l^{(j)}$.}\label{fig:complextree}
\end{figure}
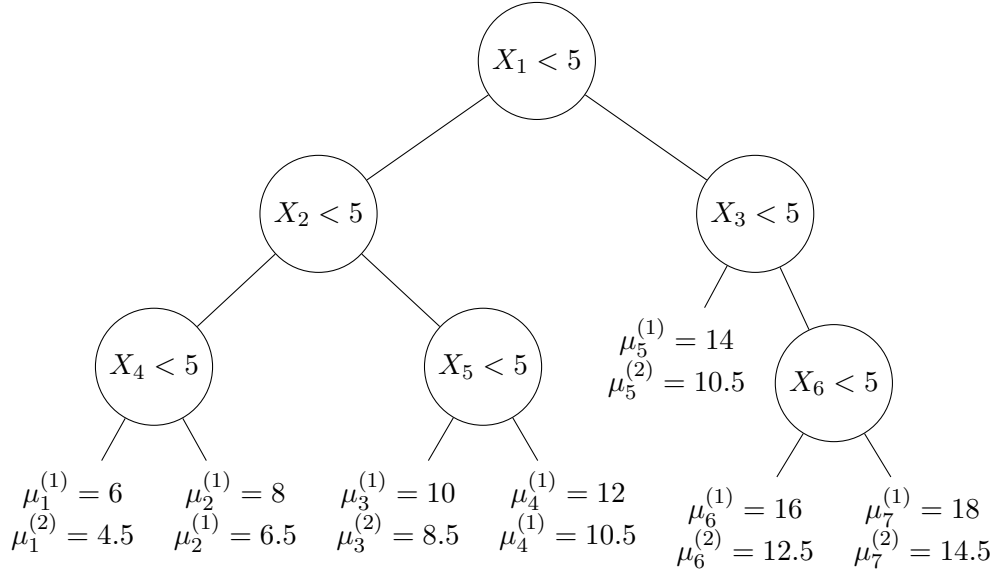

In this section, we consider a more complex tree structure as shown in Figure \ref{fig:complextree}, with seven terminal nodes. As compared to the simpler tree in Figure \ref{fig:simpletree}, this tree is deeper and has an asymmetrical structure, which makes it more difficult to estimate the fixed effects and recover the tree structure.

\begin{figure}[!htp]
	\centerline{
		\includegraphics[width=1.0\linewidth, height=0.45\textheight]{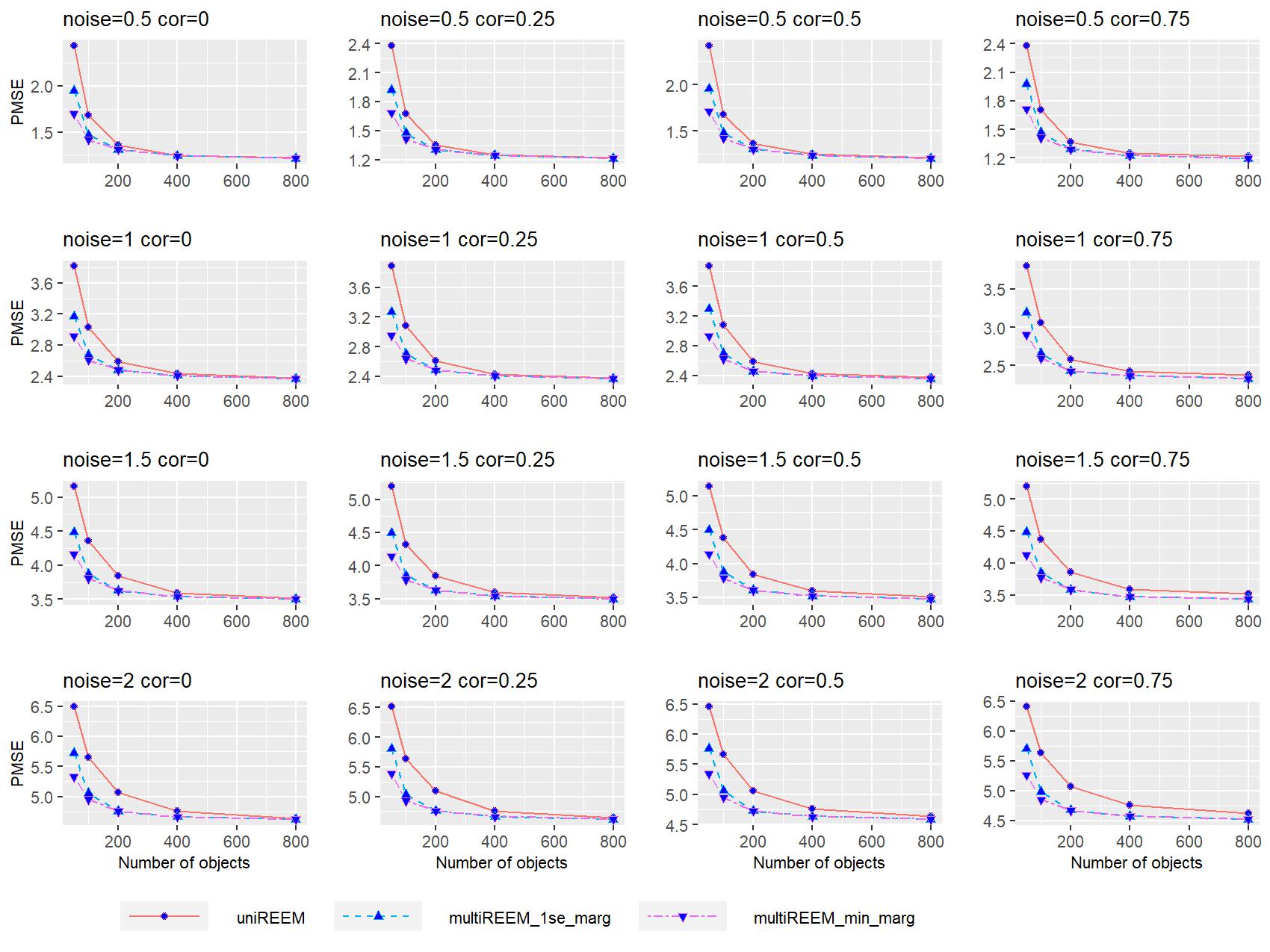}}
	\caption{PMSE in object for $T_i=5$ for the complex bivariate tree}
	\label{fig:ti5-str2-group-part1}
\end{figure}

\begin{figure}[!htp]
	\centerline{
		\includegraphics[width=1.0\linewidth, height=0.45\textheight]{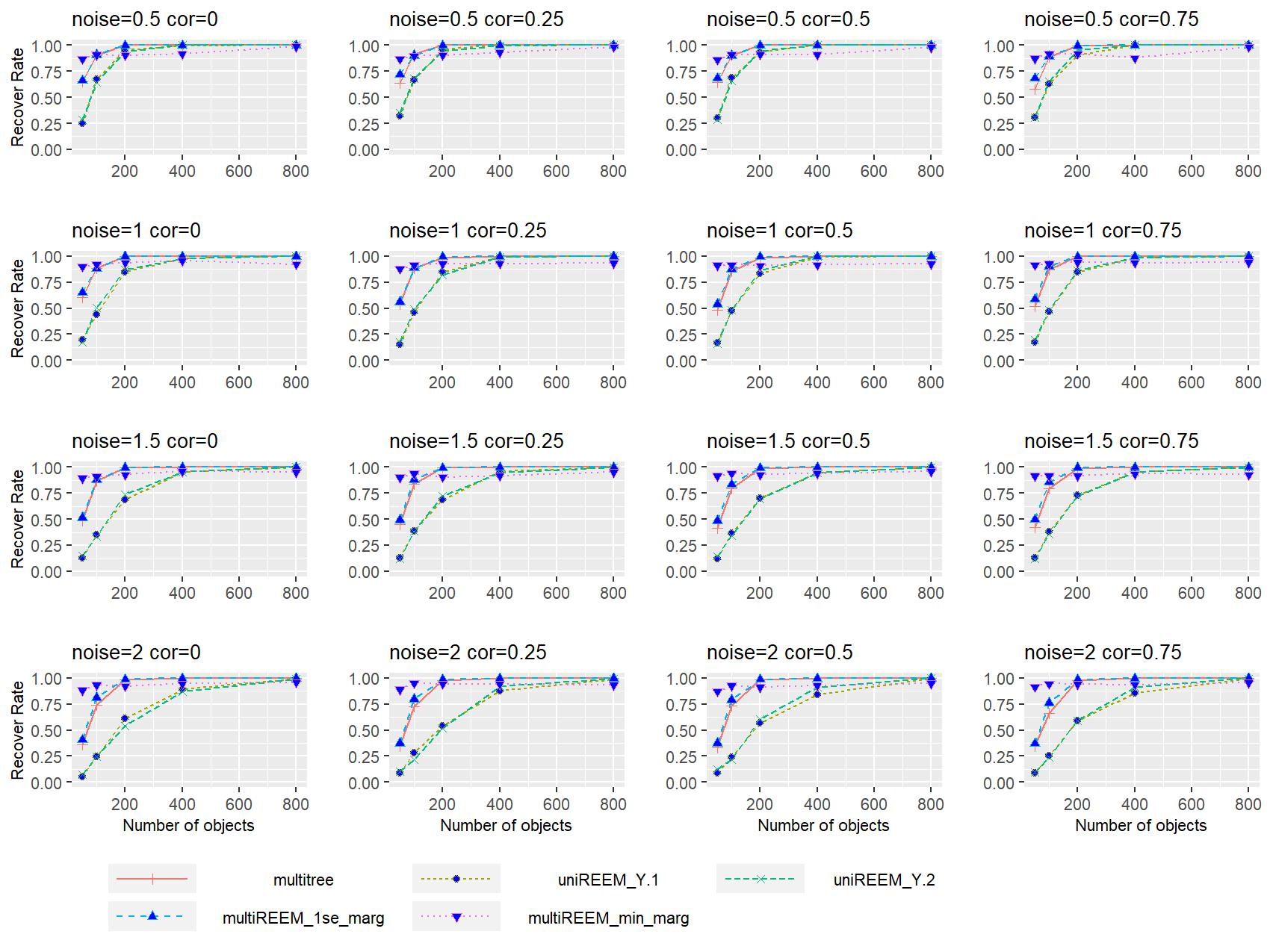}}
	\caption{Recovering rate for $T_i=5$ for the complex bivariate tree}
	\label{fig:ti5-str2-rec}
\end{figure}

\label{sec:multi-simple}

\begin{figure}[!htp]
	\centerline{
		\includegraphics[width=1.0\linewidth, height=0.45\textheight]{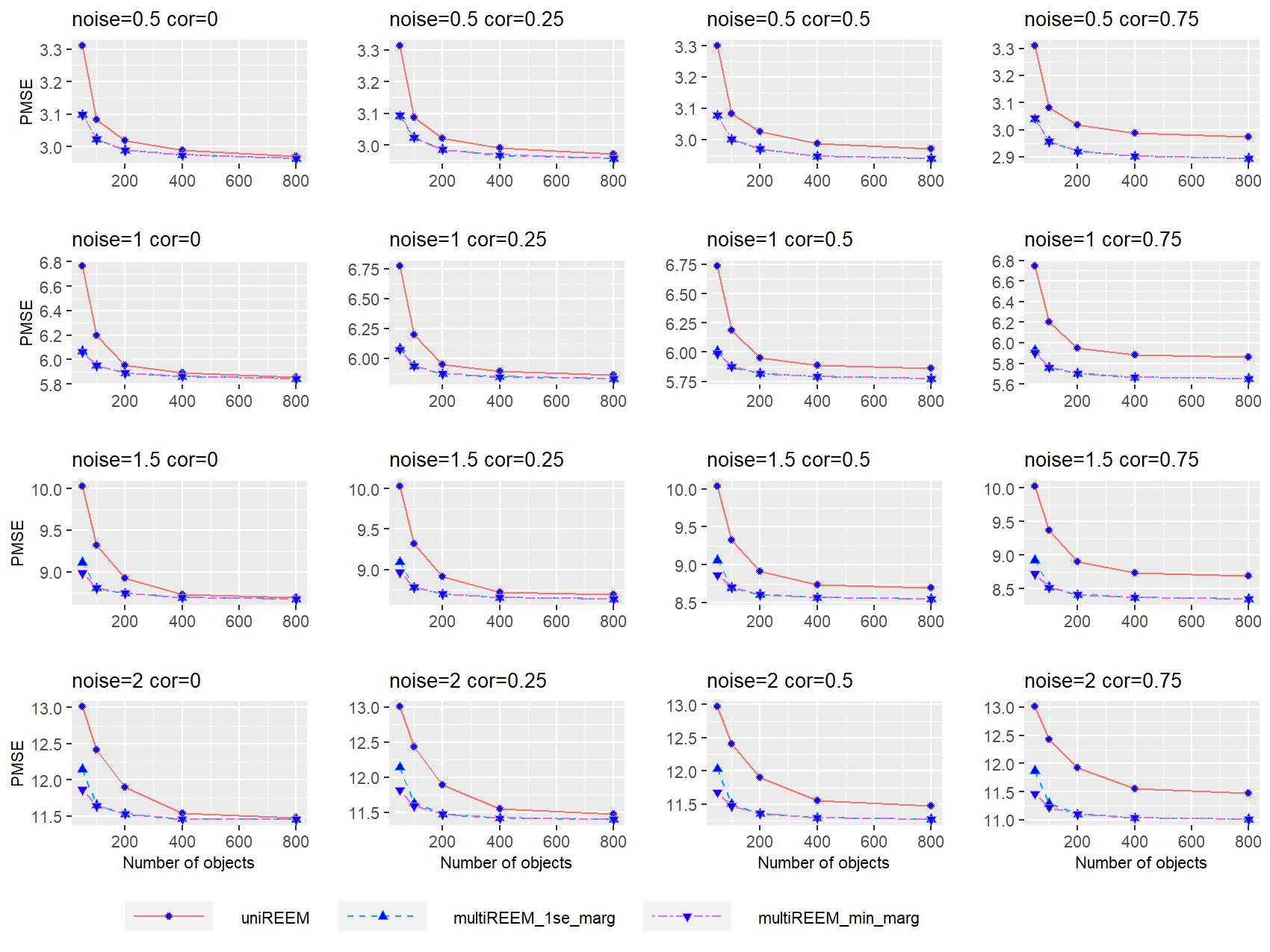}}
	\caption{PMSE in object for $T_i=5$ for multivariate tree with 5 response variables}
	\label{fig:ti5-str3-group-part1}
\end{figure}

\begin{figure}[!htp]
	\centerline{
		\includegraphics[width=1.0\linewidth, height=0.45\textheight]{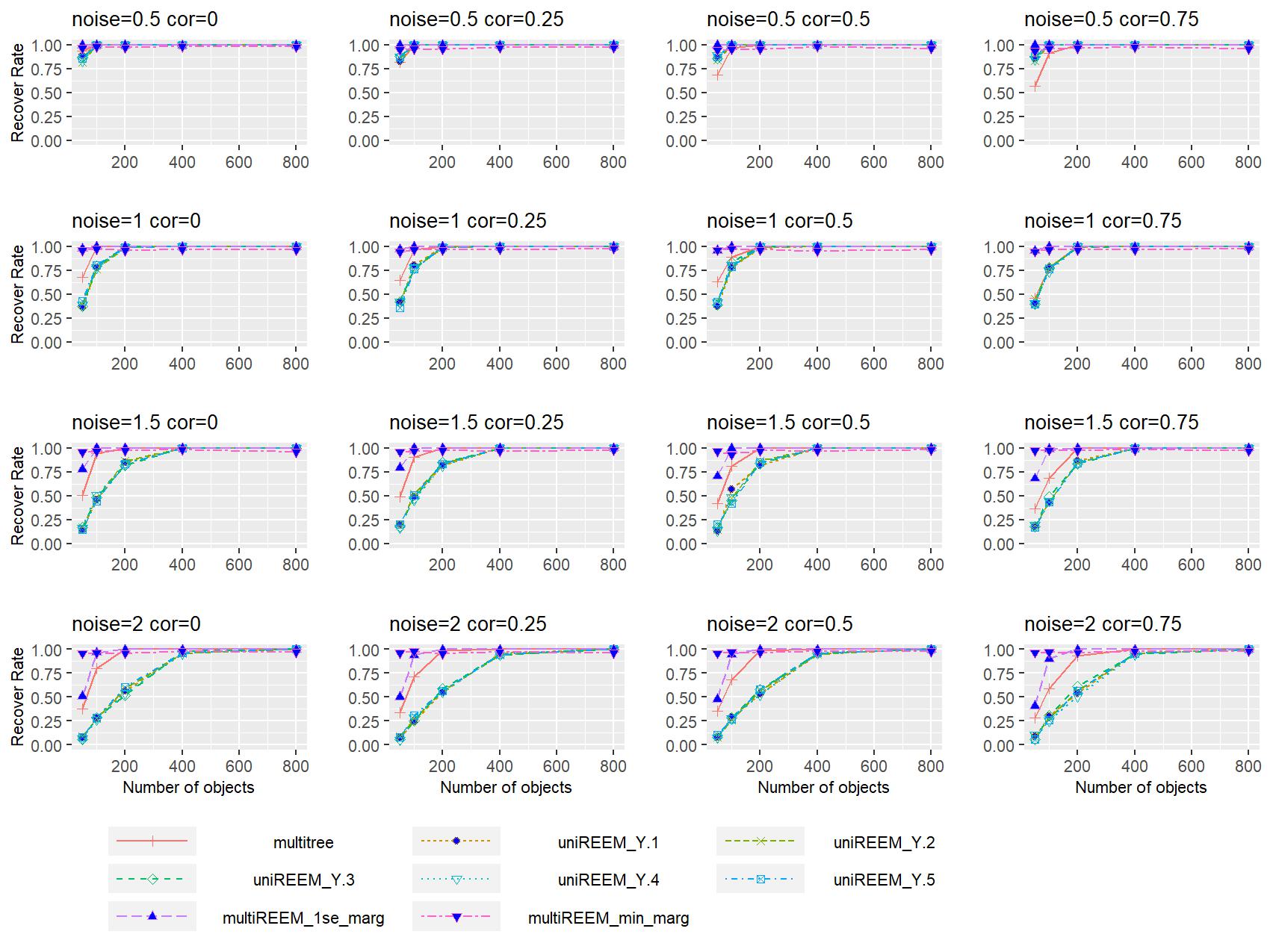}}
	\caption{Recovering rate for $T_i=5$ multivariate tree with five response variables}
	\label{fig:ti5-str3-rec}
\end{figure}

We present the object-level PMSE and the tree recovering rate for $T_i=5$ in Figure \ref{fig:ti5-str2-group-part1} and Figure \ref{fig:ti5-str2-rec}, respectively. The results for $T_i>5$ are shown in Section B of the Appendix. The patterns are similar to those seen for the simpler model, with the multivariate RE-EM tree superior in PMSE to two separate RE-EM trees. It is also clear that the simpler trees that can come from using the ``1se" rule are sometimes too simple (that is, the final trees are underfit), and result in poorer performance than using the ``min" rule, particularly for smaller sample sizes. The ``min" rule can lead to overfitting, and hence a lower recovering rate, but this clearly has relatively little effect on the PMSE compared to that of underfitting.

\subsection{More Response Variables}

In this section, we describe simulations with $J=5$ to evaluate the method with more response variables. The settings are similar to those in Section \ref{sec:setting}, only with the following tree structure: 
\begin{align*}
	&g_1=(X_1 \leq 5) \wedge (X_2 \leq 5), \bm{\mu}_1 = (10, 9, 8, 4, 6)^{\top}.\\
	&g_2=(X_1 \leq 5) \wedge (X_2 > 5), \bm{\mu}_2 = (11, 10, 9, 5, 7)^{\top}.\\
	&g_3=(X_1 > 5) \wedge (X_3 \leq 5), \bm{\mu}_3 = (12, 11, 10, 6, 8)^{\top}.\\
	&g_4=(X_1 > 5) \wedge (X_3 > 5), \bm{\mu}_4 = (13, 12, 11, 7, 9)^{\top}.
\end{align*}
where $\bm{\mu}_l=\left({\mu}^{(1)}_l,\dots,{\mu}^{(5)}_l\right)^{\top}$. For the random effect, we again take the design matrix $\bm{Z}_{it}=1$ and adjust
the covariance matrix $\bm{D}=(\sigma_{j_1j_2})_{5\times 5}$ as: $\sigma_{j_1j_2}=\sigma_B$ if $j_1\neq j_2$ and $\sigma_{j_1j_2}=1$ if $j_1=j_2$. In particular, we take $\sigma_{B}=0, .25, .5, .75$. The object-level PMSE is shown in Figure \ref{fig:ti5-str3-group-part1}, the tree recovering rate for $T_i=5$ is shown in Figure \ref{fig:ti5-str3-rec}, and results for $T_i>5$ are given in Section C of the Appendix. 


The results of the multivariate RE-EM tree with five response variables are similar to those of the bivariate tree. Moreover, with more response variables, the advantage of the multivariate RE-EM tree methods over ``uniREEM" is more pronounced, especially for higher correlation. As we hypothesized earlier, fitting a single multivariate tree apparently avoids the additional noise of fitting separate univariate longitudinal trees, and accounting for the correlation between the random effects further improves prediction performance. The tree recovering rates again verify our findings in the bivariate-tree simulations. When the sample size is small, the Multivariate RE-EM tree methods always achieve significantly higher recovering accuracy than each separate univariate RE-EM tree, and the ``1se" criterion performs worse in tree recovering compared to the ``min" criterion, which leads to a poorer prediction. However, when the sample size is large enough, the ``min" criterion may overfit the tree structure, but this does not affect the prediction. 

The primary limiting computational factor in fitting multivariate RE-EM trees is the number of response variables $J$, and in particular the fitting of the multivariate linear mixed model. All of the simulations described in this section were run using the {\tt lme} function of the R package {\tt nlme}, and further simulations indicate that computational time increases linearly with $J$. Eventually, however (at roughly $J=25$ in these computations), this becomes unworkable when using {\tt lme}. This can be overcome by using instead the {\tt MCMCglmm} function (in the same-named package) to fit the linear mixed model, although the latter function is more restrictive in the models that it can fit. Another potential approach could be to adapt the moment-based estimation scheme of \cite{perry2017fast} to the multivariate response situation.  We should note, however, that from a conceptual point of view, it seems unlikely that a single tree structure would be a very appropriate representation of the underlying random process in a situation with a large number of response variables. For this reason, we view this methodology as most useful in situations with fewer than roughly ten response variables.

\section{Real Data Example}
\label{sec:poverty}

\begin{figure}[!tp]
	\centerline{\includegraphics[width=0.89\textwidth]{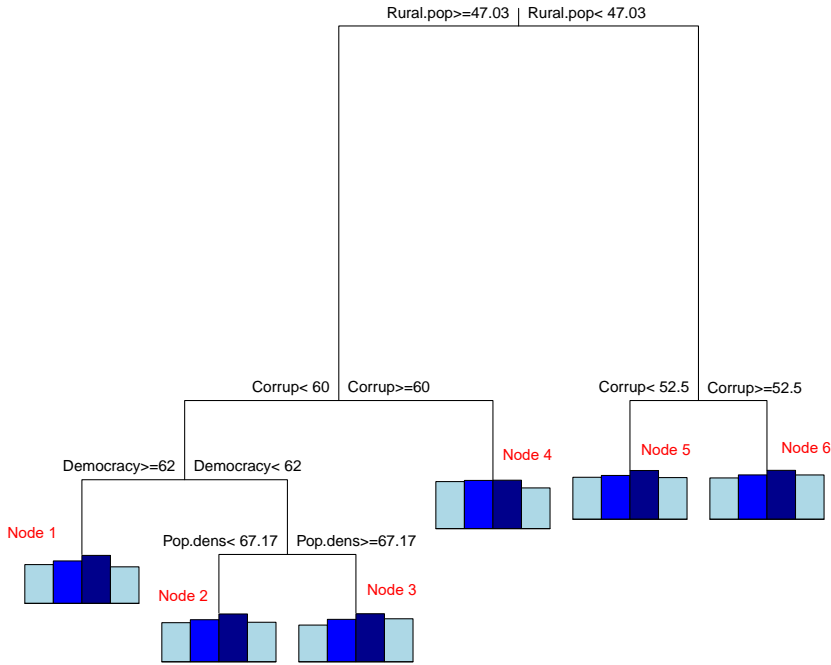}}
	\caption{Multivariate RE-EM tree of poverty data using all potential predictors. Each bar represents the average value of all country-years in that terminal node for each of the four response variables (Access to electricity, Access to safe water, Child Survival rate, and Secondary school enrollment).}
	\label{fig:tree}
\end{figure}

\begin{figure}[!tp]
	\centerline{\includegraphics[width=0.7\textwidth]{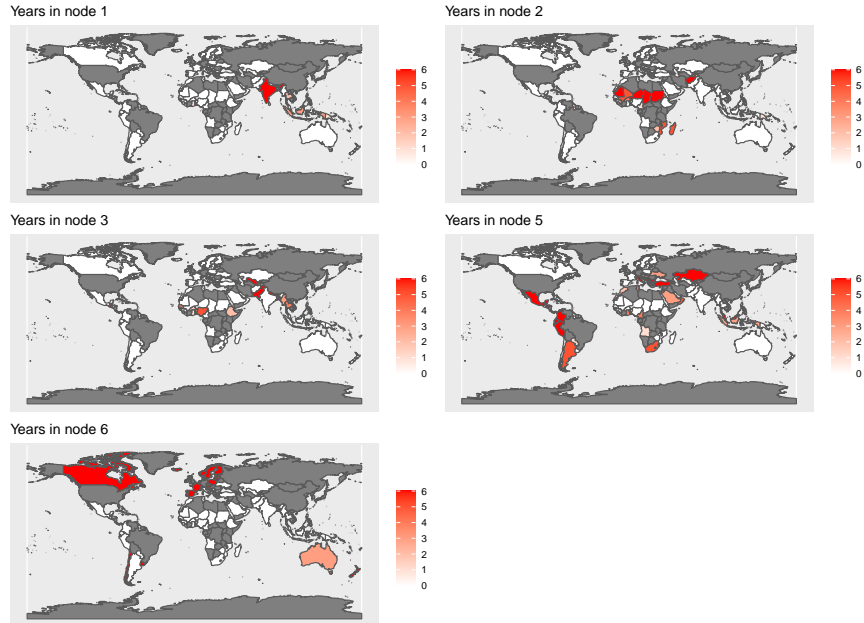}}
	\caption{Choropleth map of the number of years each country falls into each terminal node for tree using all potential predictors. Colors range from white to red as the count increases, with countries colored gray having no observations in the data set.}
	\label{fig:mapnodes}
\end{figure}

Poverty is typically measured based on income, focusing on the expenditure necessary to buy a basket of essential goods and services. Such a monetary measure does not necessarily reflect other, non-financial, aspects of well-being. The World Bank, in its 2021 document {\it A Roadmap for Countries Measuring Multidimensional Poverty\/} \citep{WorldBank} argues for the concept of multidimensional poverty, which reflects that human deprivation encompasses characteristics beyond income and consumption expenditures. Such characteristics include access to services (electricity and clean water), health care, and education. 
We examine modeling of this multidimensional character of poverty based on data from 107 countries using annual data from 2012 through 2017. We measure poverty using four response variables: percentage of population with access to electricity, percentage of population using at least basic drinking water services, child (under 5) survival rate per 1000 live births, and gross percentage secondary school enrollment, respectively. We use as potential predictors national-level variables related to factors that have been proposed as potential causes of poverty (see, for example, \url{https://en.wikipedia.org/wiki/Causes\_of\_poverty}). Among these causes of global poverty, and the variables used to measure them, are inadequate food (agriculture production index), inadequate access to health care (health expenditure per capita), inequality (Gini index), poor education (per capita education expenditures), climate change (mean surface temperature change), lack of access to livelihoods or jobs (unemployment rate), poor governance and corruption (Democracy Index (DI), Corruption Perceptions Index (CPI)), limited or poor infrastructure (rural population, trade and transport-related infrastructure index, population density), conflict (violent conflict death rate), and overpopulation (population density). Data for all of the variables are available at \cite{WorldBankdata}, with the exception of the agriculture production index \citep{agprodindex}, mean surface temperature change \citep{tempchange}, and the DI \citep{DemocIndex}. Figure \ref{fig:tree} gives the multivariate RE-EM tree for these data, with pruning based on the one-SE rule and 10-fold cross-validation, and surrogate split used to handle missing values in the predictors.

Each terminal node gives a bar plot of the average value of each response variable for all country-year pairs that fall in that node, with the bars presented in the order $\{$Access to electricity, Access to safe drinking water, Child survival rate, Secondary school enrollment$\}$. Thus, a shorter bar corresponds to a higher degree of that dimension of poverty. It should be noted, however, that those bars are in general less informative than they might appear to be, at least at the level of an individual country, since they do not reflect the predicted random effects of each response variable for that country, so we focus instead on the clustering implied by the tree: which predictors matter, in what way they matter, and how that translates to meaningful groupings of countries determined by the terminal nodes. The resultant groupings are given in the choropleth (heat) maps presented in Figure \ref{fig:mapnodes}, in which a darker color of a country corresponds to more years in which the country falls in that terminal node. 

\begin{figure}[!tp]
	\centerline{\includegraphics[width=0.89\textwidth]{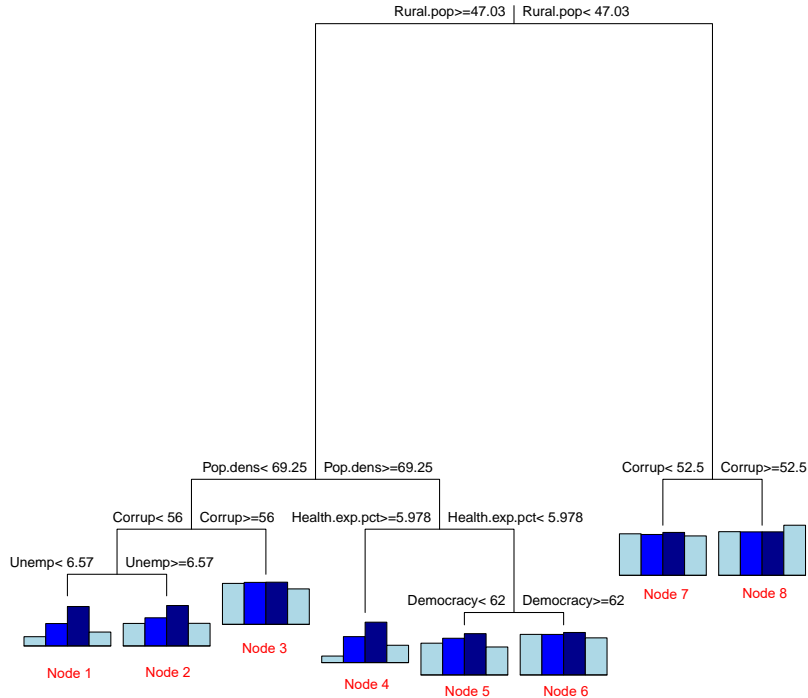}}
	\caption{MRT ({\tt mvpart}) tree of poverty data using all potential predictors.}
	\label{fig:mvparttree}
\end{figure}

The groupings produced by the tree imply familiar geographic and socioeconomic patterns, with the tree providing nonfinancial characterizations of those patterns. The country-years with higher levels of nonfinancial poverty are separated from those with lower poverty on the basis of rural population, supporting the potential importance of infrastructure. The more urban (lower poverty) country-years are then split by CPI value. The less corrupt ones form Node 6, dominated by the wealthiest countries in the sample (Canada, western Europe, Israel, Australia, and New Zealand), but also the wealthiest countries in their geographic regions (Chile, Costa Rica, and Uruguay in Central and South America, and Slovenia, Poland, and the Baltic countries in the former Communist countries of Europe), pointing to a local relative wealth effect, rather than only an absolute one. This grouping strongly parallels an ordering based on the Human Development Index (HDI) \citep{HDI}, given that all of these countries other than Costa Rica and Uruguay rank among the top 45 countries in the world in HDI. The country-years with lower rural population and higher measured corruption form Node 5. This contains the rest of South and Central America, as well as southern Europe, the Middle East, and central Asia. These geographic groupings, perhaps pointing to cultural patterns, seem more important here, as these countries range from relatively developed Italy and Greece to less developed Honduras and El Salvador, but overall this node contains a ``second tier" level of human development.

The other branch of the tree (with rural population percentage greater than 47\%) defines four groups of country-years with higher poverty rates. The first split is on CPI, which separates the six years of data for less corrupt Bhutan into Node 4, a country with notably higher access to electricity and secondary school enrollment (and thus lower nonfinancial poverty) compared to the geographically close Bangladesh, India, and Nepal (this corresponds to a distinction not revealed by the HDI, as the four countries have similar values of that measure). The corresponding map for Node 4 is not included in Figure \ref{fig:mapnodes}, since it corresponds to only one country. The remaining country-years are next split on DI, and then population density. Node 2, with lower DI and lower population density, are the poorest countries in the sample, and among the poorest in the world, concentrating in an east-west line along the southern edge of the Sahara (the Sahel), southeast Africa (Mozambique and Madagascar), and Afghanistan. Node 3 splits from these countries by having higher population density, identifying slightly wealthier countries in Africa and countries in south Asia (Bangladesh, Nepal, and Pakistan). The country-years with higher DI form Node 1, and represent a middle ground of wealth, including countries such as India, Indonesia, Mauritius, and the Philippines. Thus, the multivariate tree unsurprisingly identifies many patterns related to financial poverty, but also highlights patterns related to human development, as well as geographic (and presumably cultural) effects in nonfinancial poverty.

\begin{figure}[!tp]
	\centerline{\includegraphics[width=0.7\textwidth]{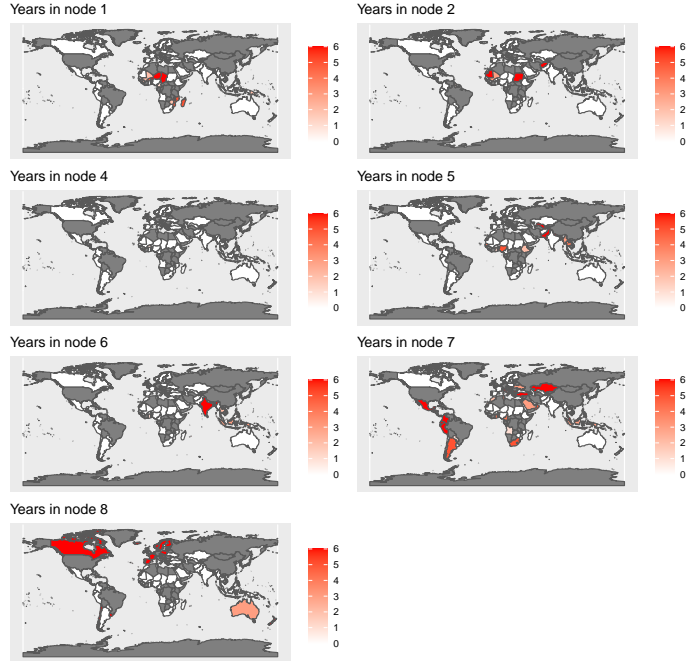}}
	\caption{Choropleth map of the number of years each country falls into each terminal node for MRT ({\tt mvpart}) tree using all potential predictors. }
	\label{fig:mvpartmapnodes}
\end{figure}

It is of course impossible to know the true underlying pattern in a real data set, but it is interesting to compare the patterns here to those implied if the longitudinal structure of the data is ignored. Figures \ref{fig:mvparttree} and \ref{fig:mvpartmapnodes} give the corresponding plots for the MRT that does not account for longitudinal structure, produced using {\tt mvpart} with pruning based on the one-SE rule and 10-fold cross-validation, and surrogate split used to handle missing values in the predictors. Unsurprisingly, the general patterns for this method are similar to those of the multivariate RE-EM tree, although the tree that does not account for longitudinal effects is more complex, having two more terminal nodes and using two additional splitting variables (this pattern of the tree that accounts for longitudinal effects being simpler was also seen in the real data examples in \cite{sela2012re}, perhaps reflecting the need to account for those effects in the fixed effects in the tree that does not include random effects). Recall that the bars given at the terminal nodes in Figures \ref{fig:tree} and \ref{fig:mvparttree} are not directly comparable to each other, since predictions for the multivariate RE-EM tree would include predicted random effects. The initial split, based on rural population percentage, is identical, as is the second split down the right side of the tree, based on CPI; this results in an identical grouping of the wealthiest countries in their regions (Nodes 5 and 6 in Figures \ref{fig:tree} and \ref{fig:mapnodes}, and Nodes 7 and 8 in Figures \ref{fig:mvparttree} and \ref{fig:mvpartmapnodes}). The observations for Bhutan are again split off by themselves (Node 4 in Figure \ref{fig:tree}, Node 3 in Figure \ref{fig:mvparttree}), although an additional split on population density is required to produce that node in the latter tree. The ``middle ground" node in the RE-EM tree (Node 1) is reproduced in the MRT (Node 6), except that Lesotho is split off into Node 2, but this again requires splitting on an extra predictor in the MRT (population density and health expenditure percentage versus CPI).

There are, however, more notable differences in the RE-EM tree and MRT for the poorest countries, which fell in Nodes 2 and 3 in Figures \ref{fig:tree} and \ref{fig:mapnodes}. These two nodes are split into four nodes in the MRT, with the poorest countries (Node 2) split into Nodes 1 and 2 in Figures \ref{fig:mvparttree} and \ref{fig:mvpartmapnodes}. Note that this breaks up the appealing grouping of the Sahel countries from the Atlantic Ocean to the Red Sea that occurs in Node 2 of the RE-EM tree. The next poorest countries (Node 3 of Figures \ref{fig:tree} and \ref{fig:mapnodes}) are split into Nodes 4 and 5 of Figures \ref{fig:mvparttree} and \ref{fig:mvpartmapnodes}. the observations of Node 4 are those of the geographically diverse African countries Burundi, Malawi, Rwanda, and Sierra Leone. There is nothing apparently very different about those countries compared to the ones in Node 5, so it is not clear why the MRT makes this split.

The multivariate trees are not directly comparable to constructing separate trees for each response variable, since the separate trees will in general be different from each other, and that is certainly the case here. Separate univariate RE-EM trees (not presented here) range in complexity from four terminal nodes (access to drinking water and survival rate) to six terminal nodes (access to electricity) to nine terminal nodes (school enrollment), which of course defeats the purpose of constructing a single coherent set of groups. All four trees split first on rural population, and all include population density as a split variable, but otherwise they include different predictors from each other, including two (temperature change and Gini index) that did not appear in either of the multivariate trees.

\section{Conclusion}
\label{sec:concl}

In this paper we have described the multivariate RE-EM tree, a generalization of the univariate RE-EM tree to multivariate response data, including generalization to non-Gaussian distributions. Longitudinal trees for univariate response data have been generalized in various ways, and potential future work would be to explore such generalizations for multivariate responses. Possibilities include  random forests \citep{capitaine2021random, hajjem2014mixed} and more general model fits at terminal nodes than a single location summary statistic for responses \citep{fokkema2018detecting, stegmann2018recursive, wei2020precision}.

An R package to implement the multivariate RE-EM tree, {\tt multiREEMtree}, is available at \url{https://github.com/WenboJing1998/Multivariate-RE-EM-tree}. The code to implement the simulations and data analysis can also be found there. The simulations were run on R 4.0.3 using the NYU High Performance Computing cluster (NYU HPC), and the data analysis was run on R 4.1.1 under Windows 10 Pro.  

\section*{Acknowledgment}

We would like to thank an anonymous referee for helpful comments that helped improve the paper.

\section*{Statements and Declarations}

\textbf{Conflicts of interest} The authors have no conflicts of interest to declare that are relevant to the content of this article.

\bibliography{MultiREEM}

\newpage
\appendix

\renewcommand\thesection{\Alph{section}}
\setcounter{section}{0}

\section{Additional Results for Simpler Bivariate Tree}
\label{sec:SM:EMSE}

\subsection{Object-Level PMSE for All of the Competitor Methods}

\begin{figure}[!htp]
	\centerline{
		\includegraphics[width=1.0\linewidth, height=0.4\textheight]{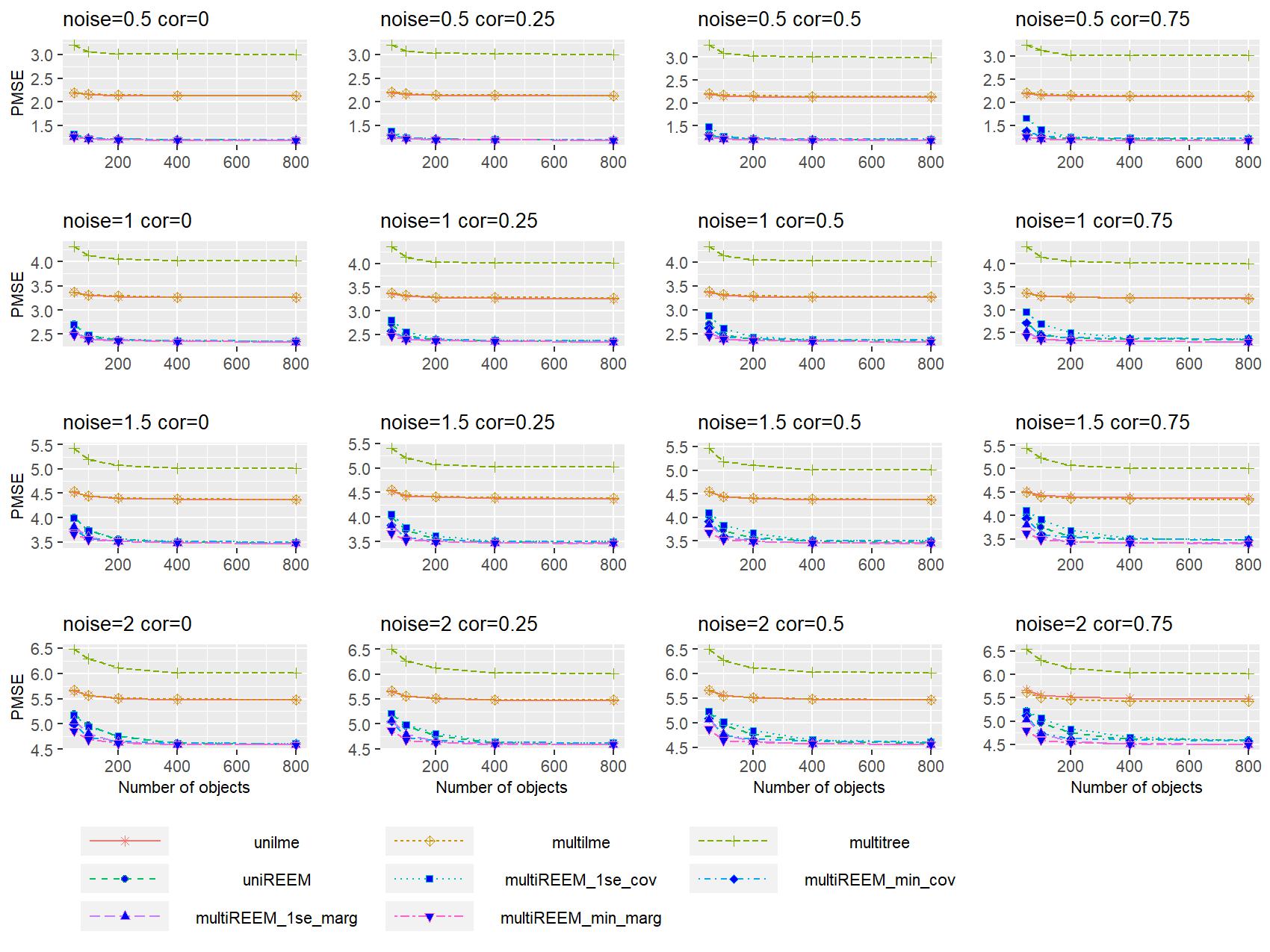}}
	\caption{PMSE in object for $T_i = 5$ for the simpler bivariate tree structure}
	\label{fig:SM:ti5-str1-group}
\end{figure}

\begin{figure}[!htp]
	\centerline{
		\includegraphics[width=1.0\linewidth, height=0.4\textheight]{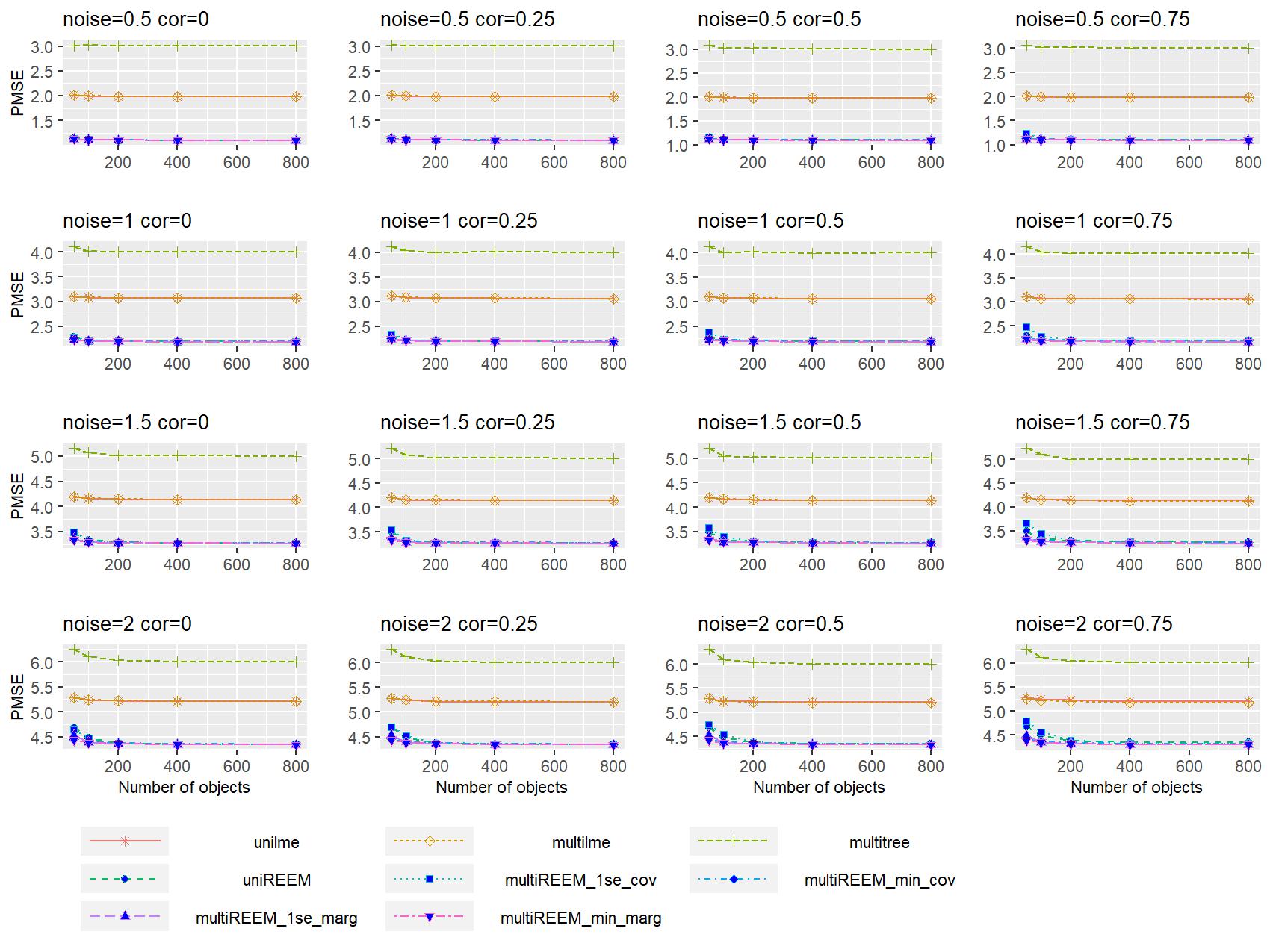}}
	\caption{PMSE in object for $T_i = 10$ for the simpler bivariate tree structure}
	\label{fig:SM:ti10-str1-group}
\end{figure}

\begin{figure}[!htp]
	\centerline{
		\includegraphics[width=1.0\linewidth, height=0.4\textheight]{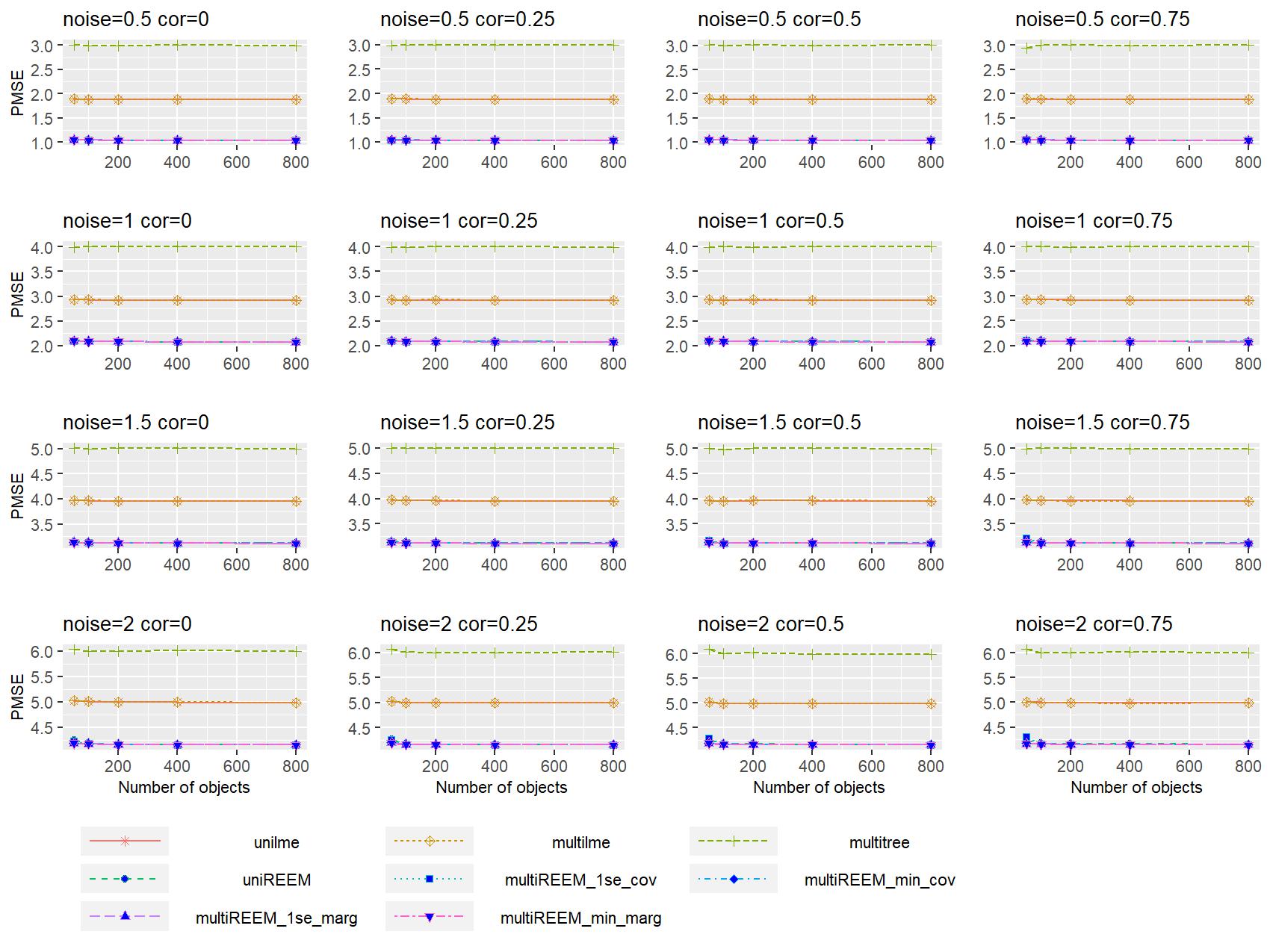}}
	\caption{PMSE in object for $T_i = 25$ for the simpler bivariate tree structure}
	\label{fig:SM:ti25-str1-group}
\end{figure}

\begin{figure}[!htp]
	\centerline{
		\includegraphics[width=1.0\linewidth, height=0.4\textheight]{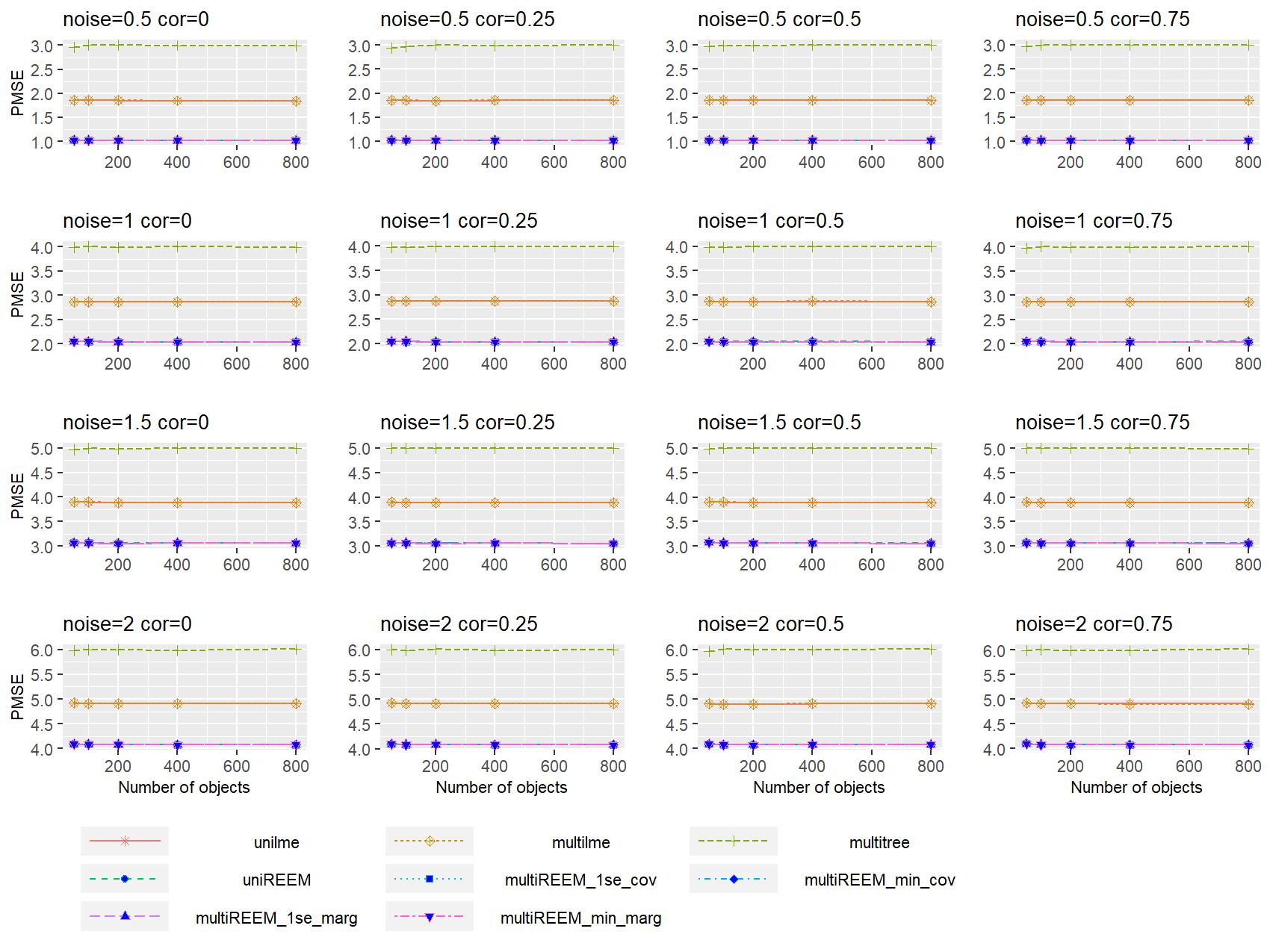}}
	\caption{PMSE in object for $T_i = 50$ for the simpler bivariate tree structure}
	\label{fig:SM:ti50-str1-group}
\end{figure}

We present the full results for all of the methods in Figures \ref{fig:SM:ti5-str1-group}--\ref{fig:SM:ti50-str1-group} here for completeness. It is clear from the results that the RE-EM methods always outperform the linear methods and the non-longitudinal tree method.

\subsection{Fixed Effects Estimation and Random Effects Prediction}

\begin{figure}[!htp]
	\centerline{
		\includegraphics[width=1.0\linewidth, height=0.4\textheight]{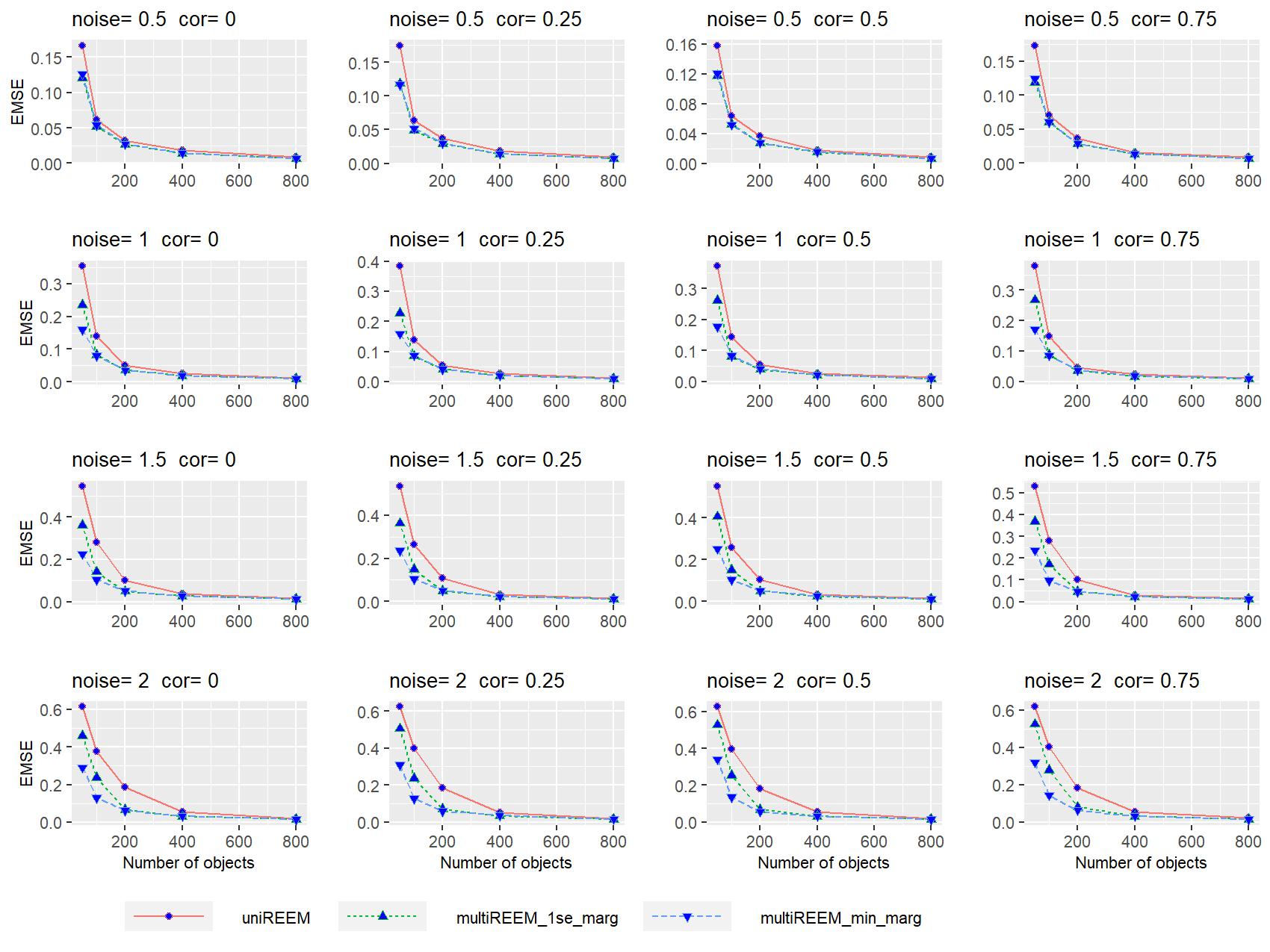}}
	\caption{Estimation Mean Squared Error for fixed effects with $T_i=5$}
	\label{fig:SM:ti5-str1-fixed}
\end{figure}

\begin{figure}[!htp]
	\centering
	\includegraphics[width=1.0\linewidth, height=0.4\textheight]{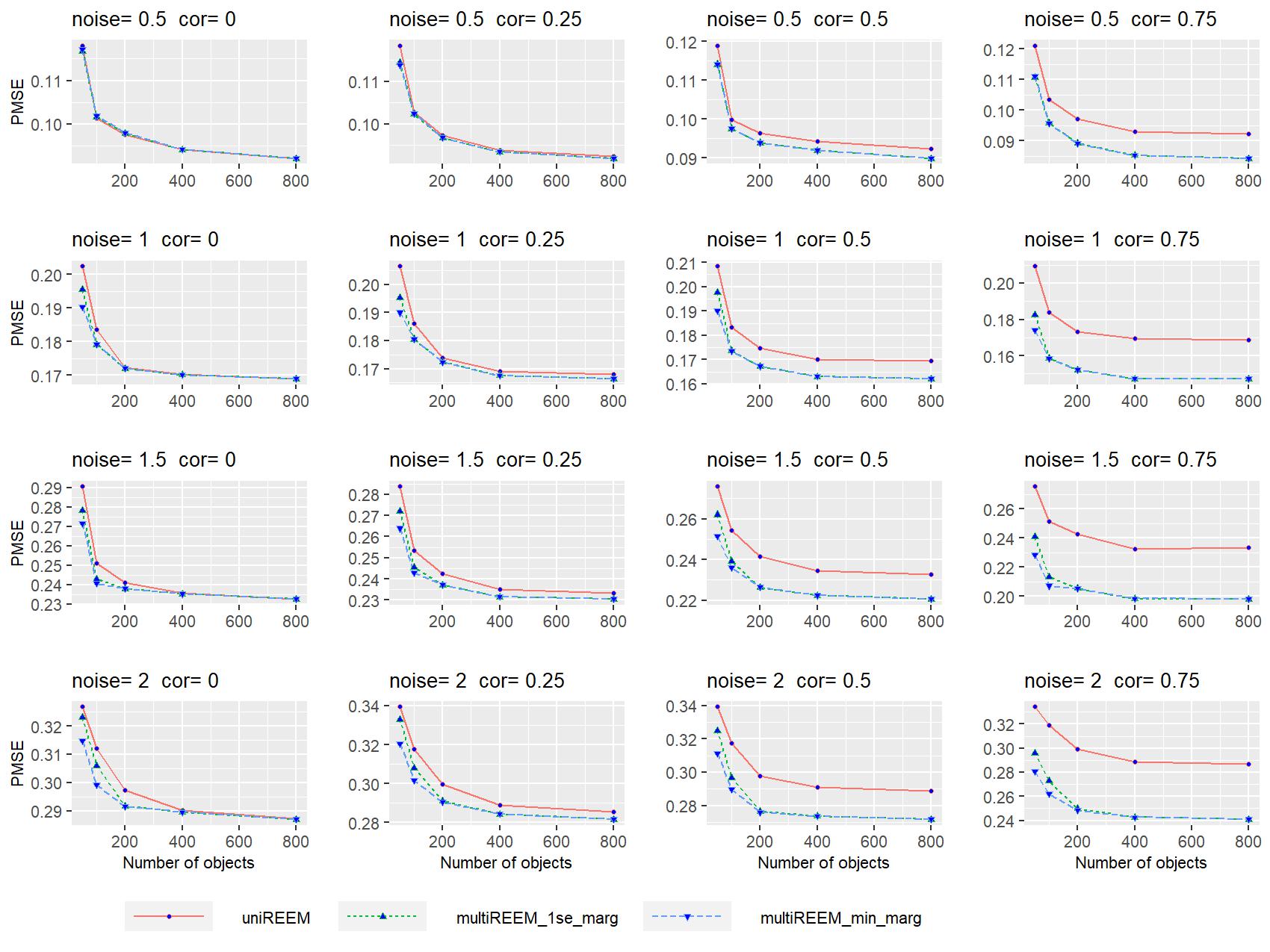}
	\caption{Prediction Mean Squared Error for random effects with $T_i=5$}
	\label{fig:SM:ti5-str1-RE}
\end{figure}

\begin{figure}[!htp]
	\centering
	\includegraphics[width=1.0\linewidth, height=0.4\textheight]{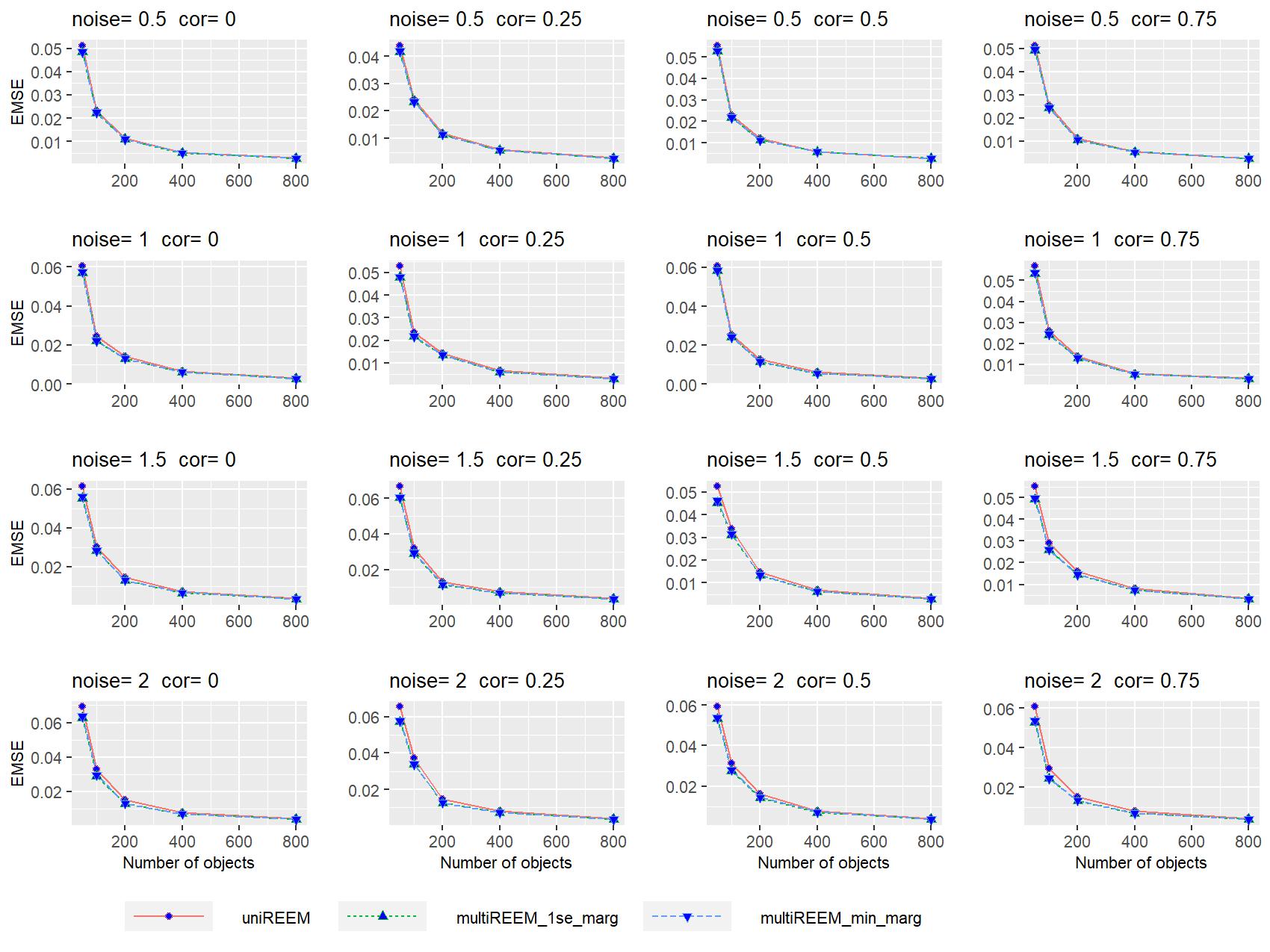}
	\caption{Estimation Mean Squared Error for fixed effects with $T_i=50$}
	\label{fig:SM:ti50-str1-fixed}
\end{figure}

\begin{figure}[!htp]
	\centering
	\includegraphics[width=1.0\linewidth, height=0.4\textheight]{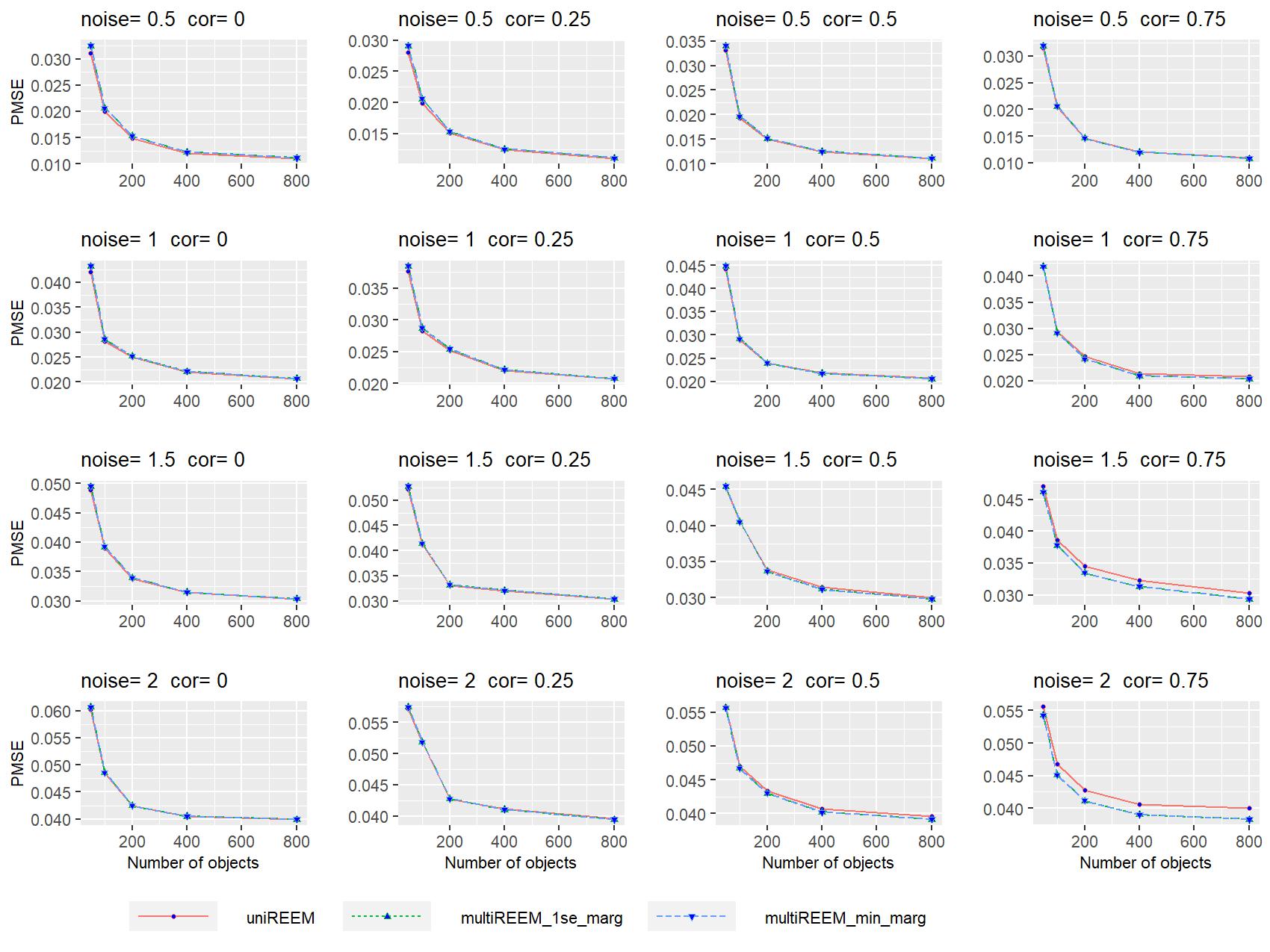}
	\caption{Prediction Mean Squared Error for random effects with $T_i=50$}
	\label{fig:SM:ti50-str1-RE}
\end{figure}

Figures \ref{fig:SM:ti5-str1-fixed}--\ref{fig:SM:ti50-str1-RE} summarize the \textit{estimation mean squared error} (EMSE) for the fixed effects and the \textit{prediction mean squared error} (PMSE) for the random effects in the simpler bivariate tree setting.  The Multivariate RE-EM tree method always estimates the fixed effects slightly better than ``uniREEM," although their recovering rates are almost the same. This indicates that separate univariate RE-EM trees are able to recover the tree structure correctly but cannot estimate the true values on each terminal node as well as the Multivariate RE-EM tree. For the random effects, the Multivariate RE-EM tree also performs no worse than ``uniREEM," and it significantly outperforms ``uniREEM" as the noise or the correlation increases.

%
%
%
%
%

\begin{figure}[!htp]
	\centering
	\includegraphics[width=1.0\linewidth, height=0.4\textheight]{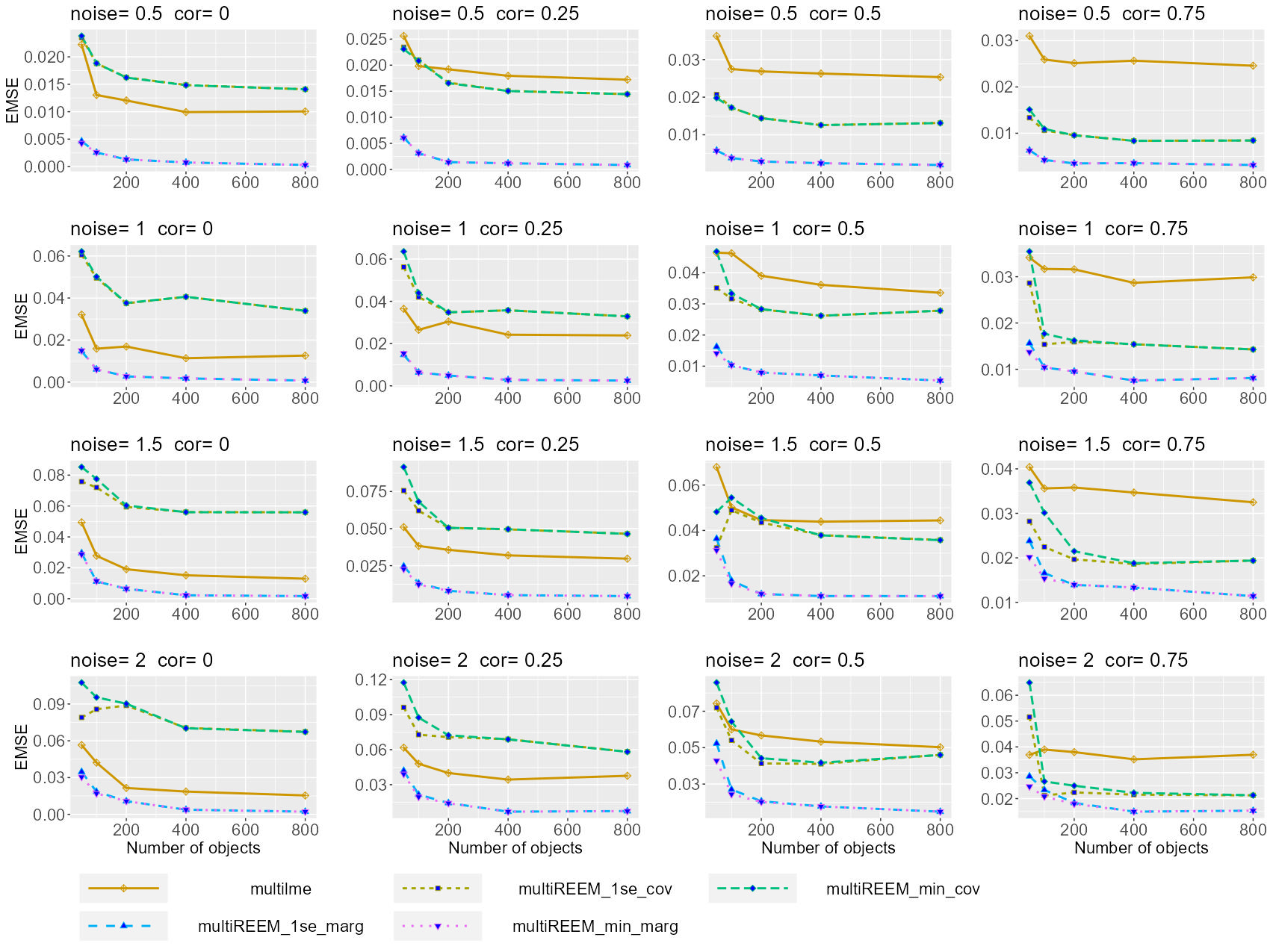}
	\caption{Estimation Mean Squared Error for the correlation $\sigma_{12}$ of the random effects with $T_i=5$}
	\label{fig:SM:ti5-str1-sigmab}
\end{figure}
\begin{figure}[!htp]
	\centering
	\includegraphics[width=1.0\linewidth, height=0.4\textheight]{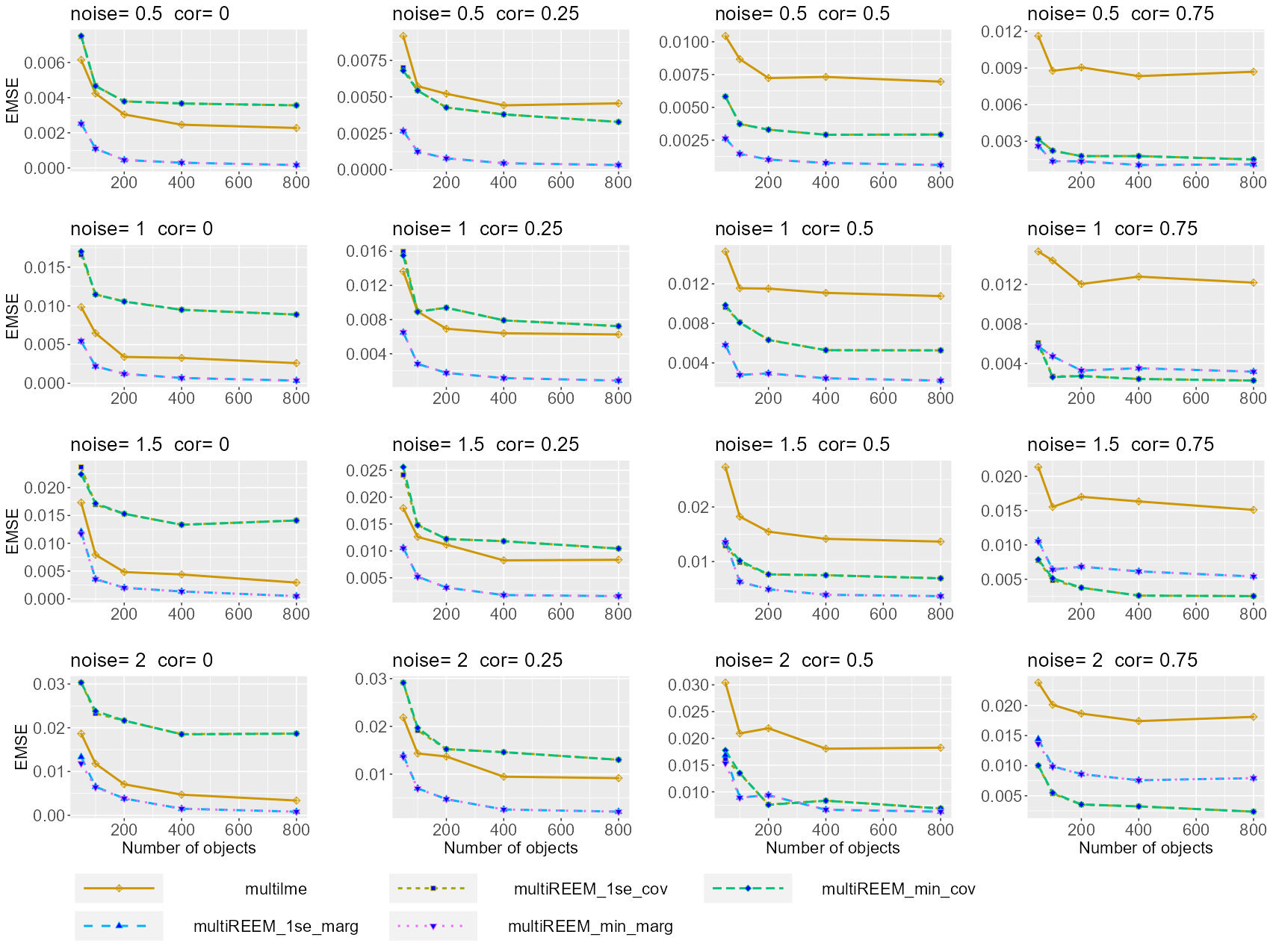}
	\caption{Estimation Mean Squared Error for the correlation $\sigma_{12}$ of the random effects with $T_i=10$}
	\label{fig:SM:ti10-str1-sigmab}
\end{figure}
\begin{figure}[!htp]
	\centering
	\includegraphics[width=1.0\linewidth, height=0.4\textheight]{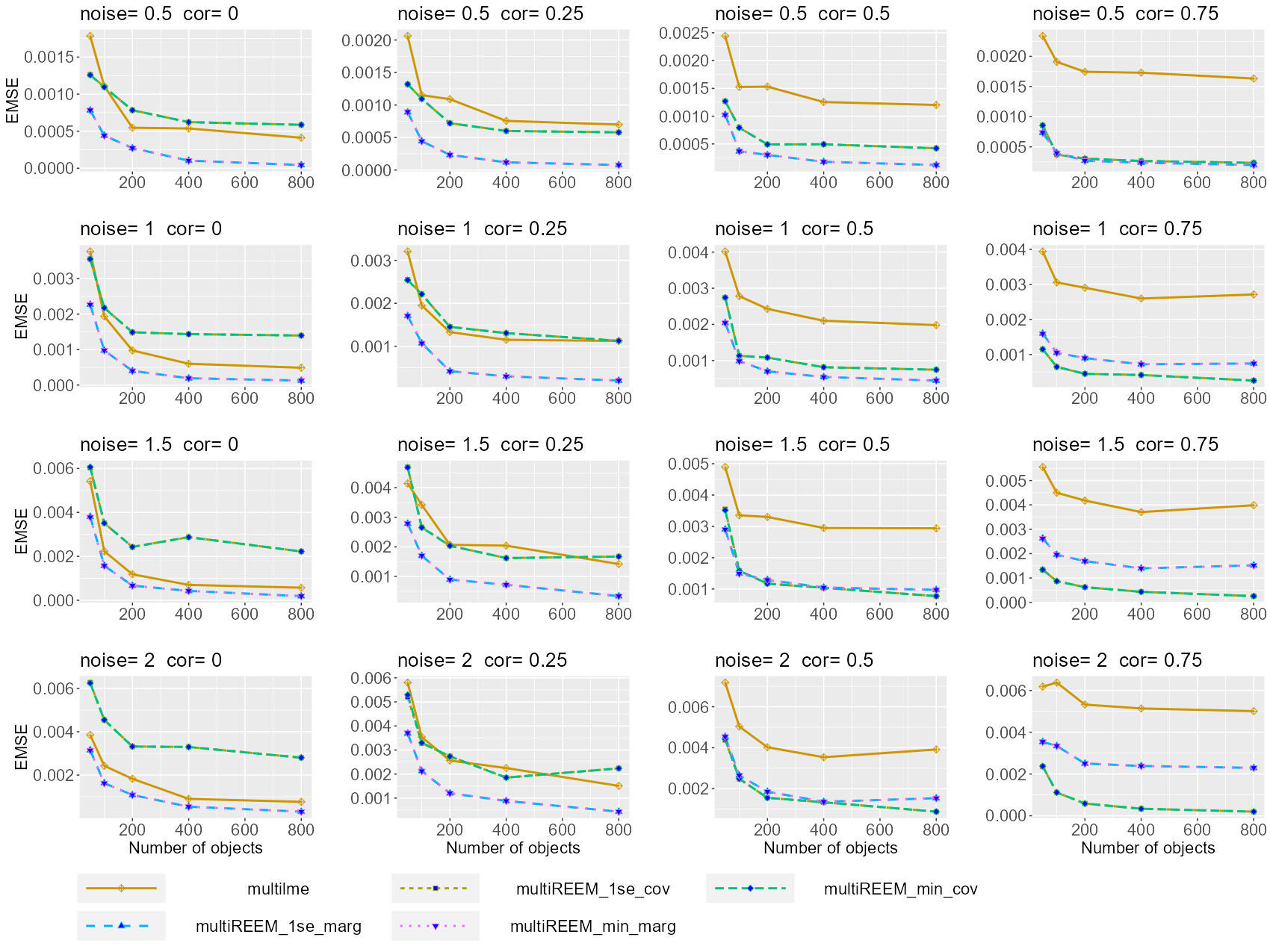}
	\caption{Estimation Mean Squared Error for the correlation $\sigma_{12}$ of the random effects with $T_i=25$}
	\label{fig:SM:ti25-str1-sigmab}
\end{figure}
\begin{figure}[!htp]
	\centering
	\includegraphics[width=1.0\linewidth, height=0.4\textheight]{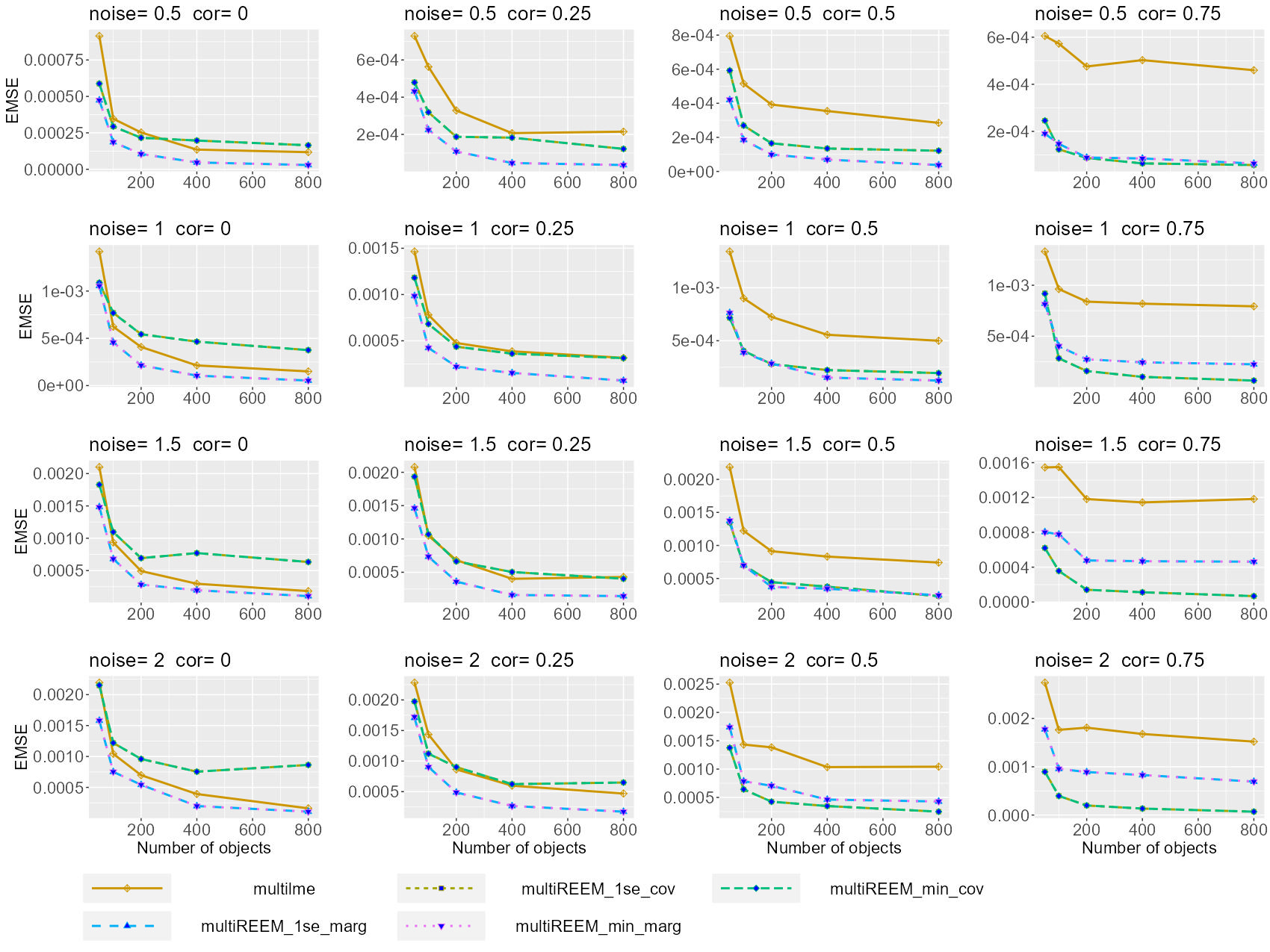}
	\caption{Estimation Mean Squared Error for the correlation $\sigma_{12}$ of the random effects with $T_i=50$}
	\label{fig:SM:ti50-str1-sigmab}
\end{figure}

Additionally, we provide the EMSEs for estimating the correlation $\sigma_{12}$ of the random effects in Figures \ref{fig:SM:ti5-str1-sigmab}--\ref{fig:SM:ti50-str1-sigmab}. The competitors include the ``multilme" method and all versions of ``multiREEMtree" methods. The true value of $\sigma_{12}$ is given by ``cor" above each subfigure, and the average values of $(\widehat\sigma_{12}-\sigma_{12})^2$ over 100 independent runs are plotted. The estimation errors of Multivariate RE-EM tree with ``marginal" standardization are always significantly lower than ``multilme," and the Multivariate RE-EM tree with ``covariance" standardization performs better when the true correlation is larger.

\begin{figure}[!htp]
	\centering
	\includegraphics[width=1.0\linewidth, height=0.4\textheight]{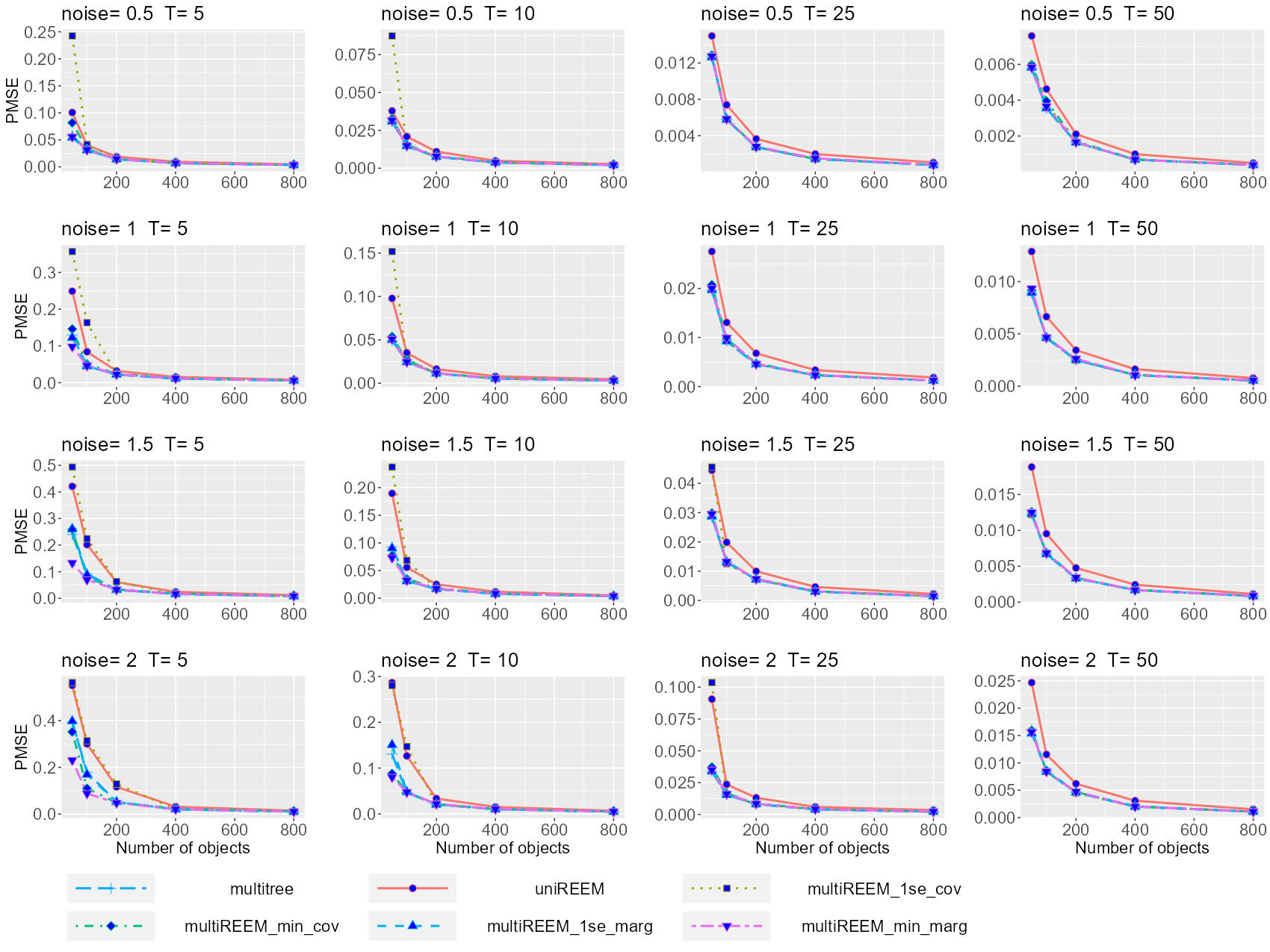}
	\caption{EMSE of the fixed effect for the simulations with no random effect}
	\label{fig:SM:norandom-fixed}
\end{figure}

\begin{figure}[!htp]
	\centering
	\includegraphics[width=1.0\linewidth, height=0.4\textheight]{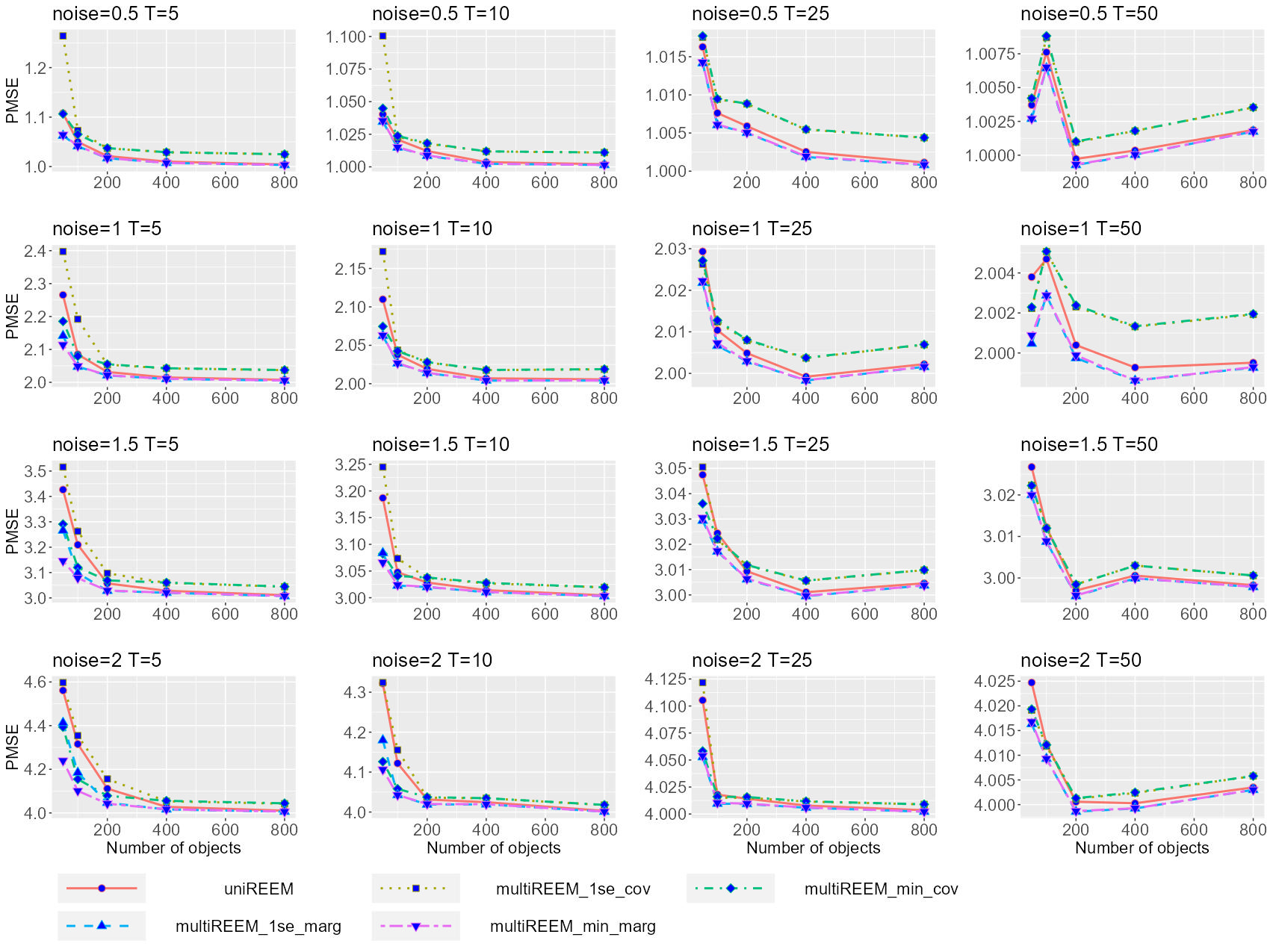}
	\caption{PMSE in object for the simulations with no random effect}
	\label{fig:SM:norandom-group}
\end{figure}

\subsection{No-Random-Effect Simulations}

In this section, we show the results of simulations with no random effects, i.e., $\bm{B}=0$. Figures \ref{fig:SM:norandom-fixed} and \ref{fig:SM:norandom-group} present the EMSE of the fixed effect and the PMSE for object-level prediction, where the competitors are all of the tree-based methods. The MRT (``multitree") method and the multivariate RE-EM trees with ``marginal" standardization always outperform other method, with no meaningful difference between them. This verifies that the proposed ``multiREEMtree" algorithm works as well as an MRT in no-random-effect cases.

\section{Additional Results for the Complex Bivariate Tree Structure}

\begin{figure}[!htp]
	\centering
	\includegraphics[width=1.0\linewidth, height=0.4\textheight]{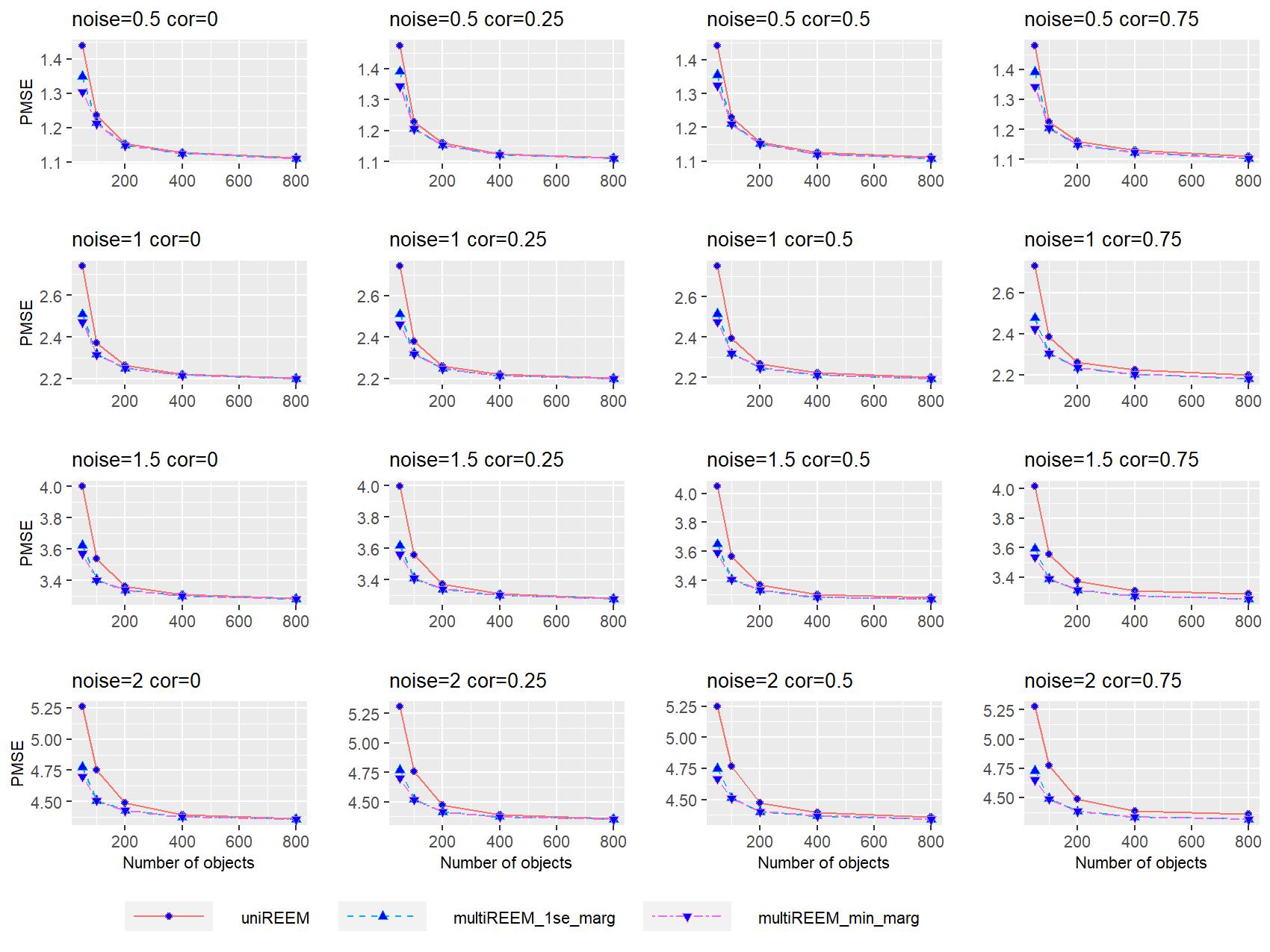}
	\caption{PMSE in object for $T_i=10$ for the complex bivariate tree}
	\label{fig:ti10-str2-group-part1}
\end{figure}

\begin{figure}[!htp]
	\centering
	\includegraphics[width=1.0\linewidth, height=0.4\textheight]{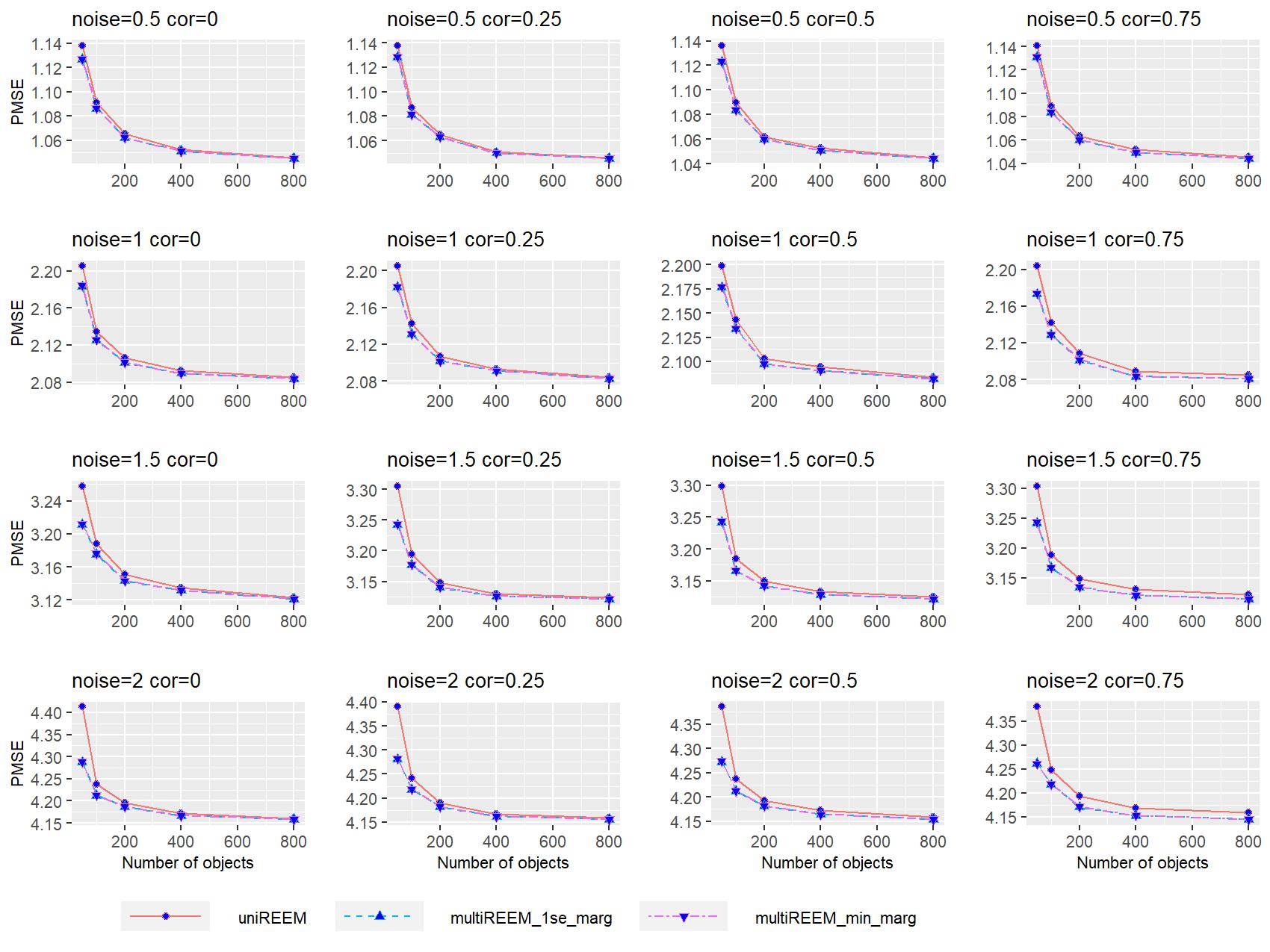}
	\caption{PMSE in object for $T_i=25$ for the complex bivariate tree}
	\label{fig:ti25-str2-group-part1}
\end{figure}

\begin{figure}[!htp]
	\centering
	\includegraphics[width=1.0\linewidth, height=0.4\textheight]{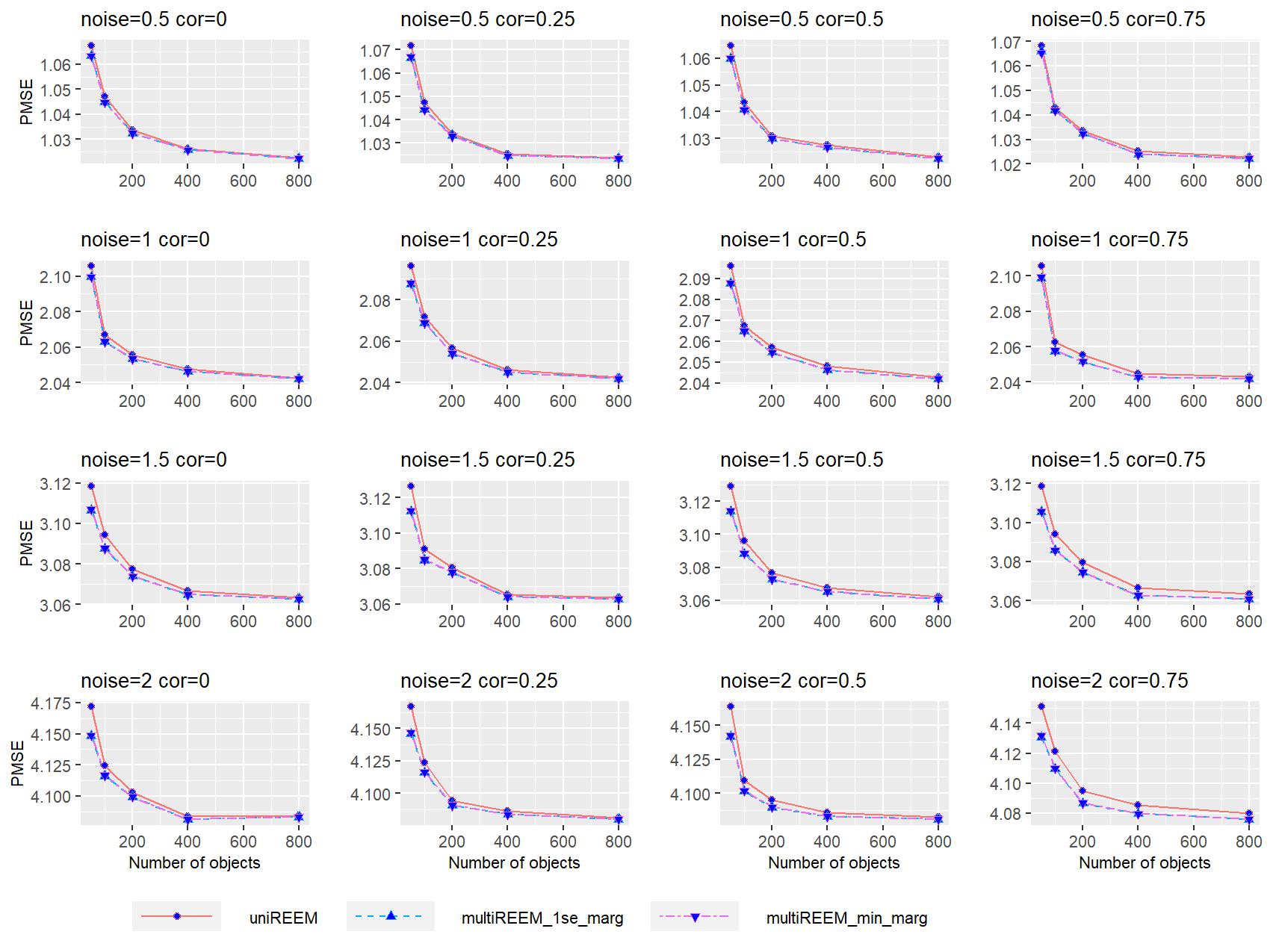}
	\caption{PMSE in object for $T_i=50$ for the complex bivariate tree}
	\label{fig:ti50-str2-group-part1}
\end{figure}

\begin{figure}[!htp]
	\centering
	\includegraphics[width=1.0\linewidth, height=0.4\textheight]{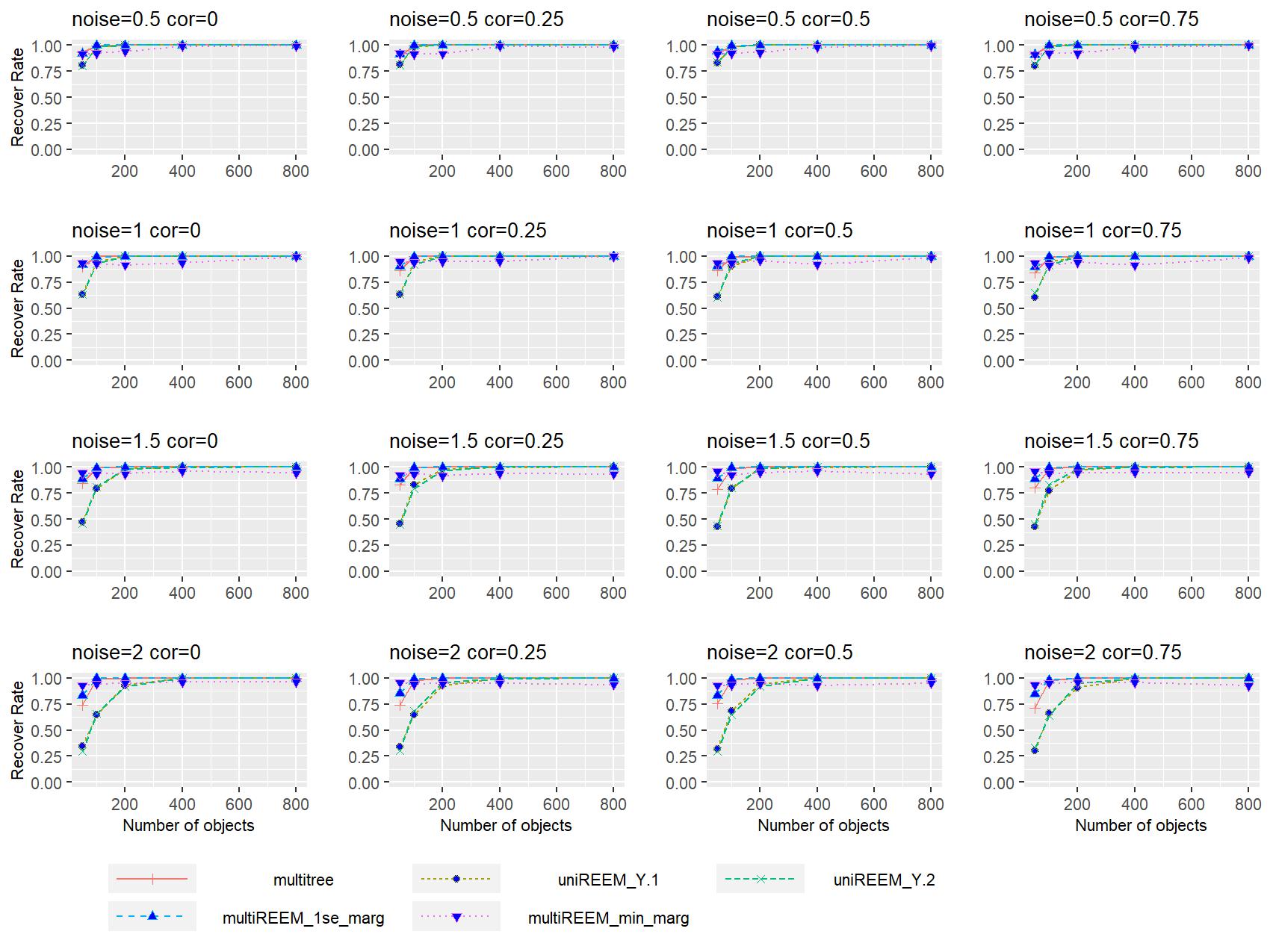}
	\caption{Recovering rate for $T_i=10$ for the complex bivariate tree}
	\label{fig:ti10-str2-rec}
\end{figure}

\begin{figure}[!htp]
	\centering
	\includegraphics[width=1.0\linewidth, height=0.4\textheight]{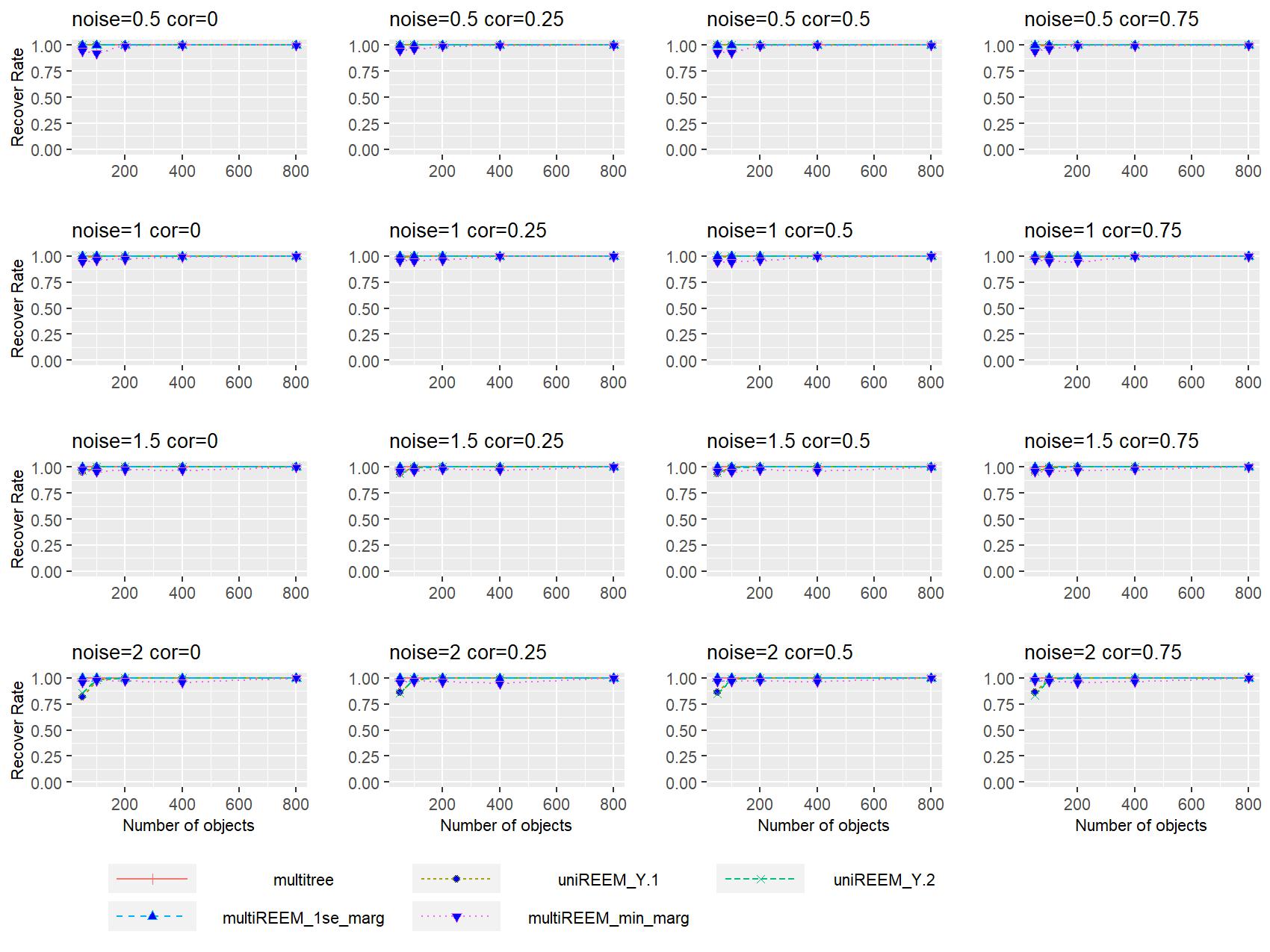}
	\caption{Recovering rate for $T_i=25$ for the complex bivariate tree}
	\label{fig:ti25-str2-rec}
\end{figure}

\begin{figure}[!htp]
	\centering
	\includegraphics[width=1.0\linewidth, height=0.4\textheight]{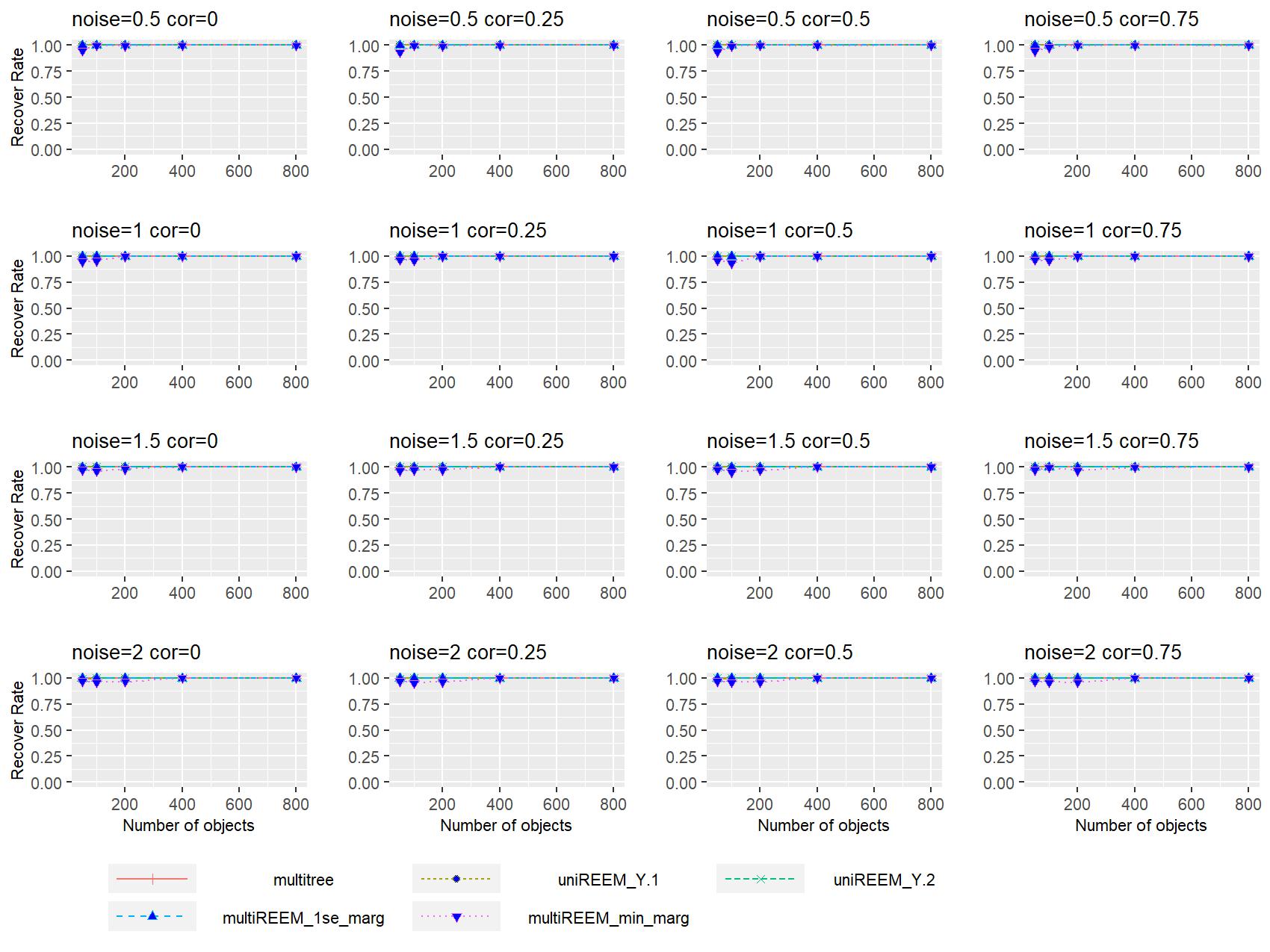}
	\caption{Recovering rate for $T_i=50$ for the complex bivariate tree}
	\label{fig:ti50-str2-rec}
\end{figure}

The PMSEs and the tree recovering rates for $T_i=10, 25, 50$ with the complex bivariate tree structure are shown in Figures \ref{fig:ti10-str2-group-part1}--\ref{fig:ti50-str2-rec}.



\section{Additional Results for Multivariate Tree with Five Response Variables}

\begin{figure}[!htp]
	\centering
	\includegraphics[width=1.0\linewidth, height=0.4\textheight]{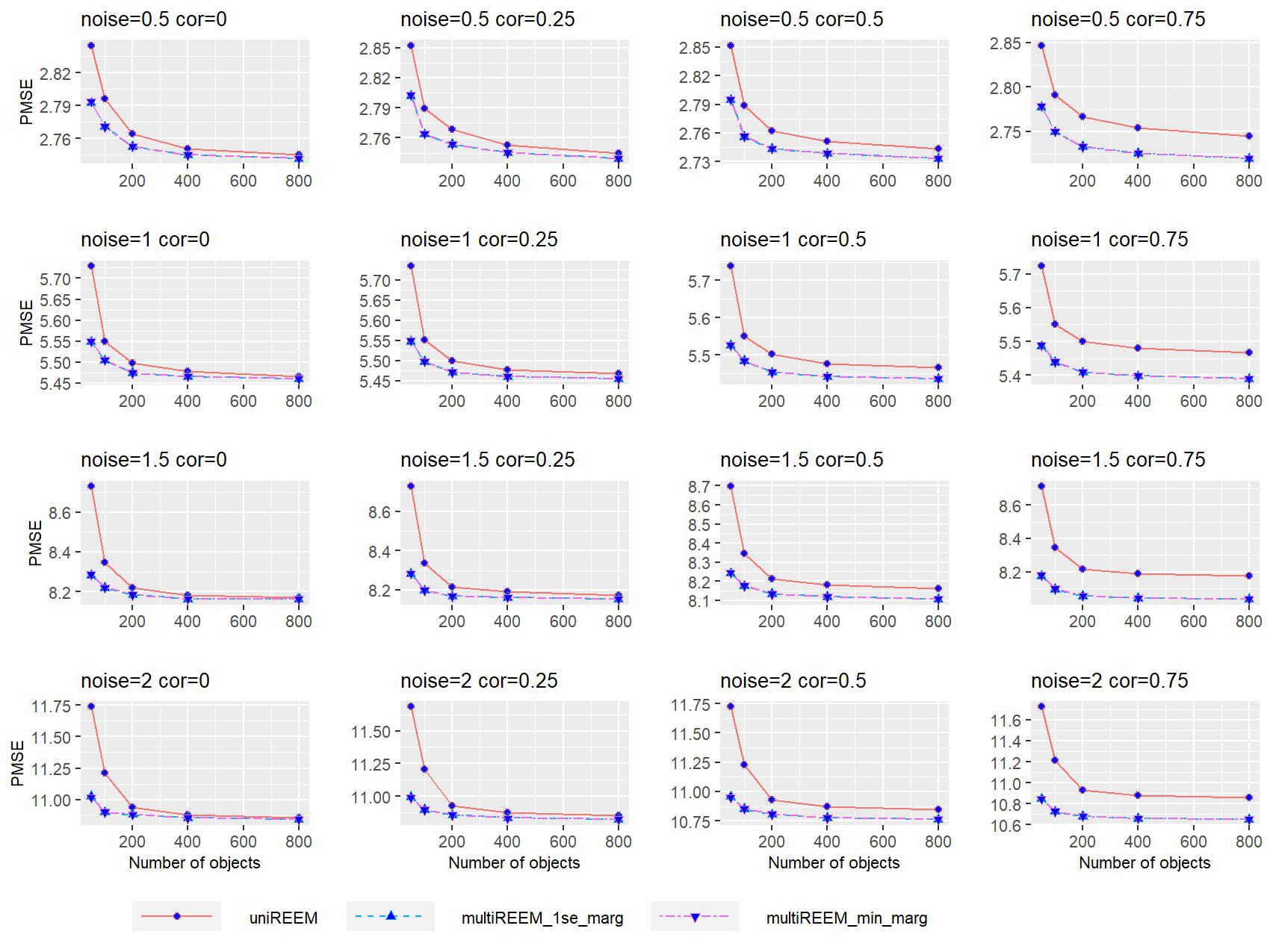}
	\caption{PMSE in object for $T_i=10$ for multivariate tree with five response variables}
	\label{fig:ti10-str3-group-part1}
\end{figure}

\begin{figure}[!htp]
	\centering
	\includegraphics[width=1.0\linewidth, height=0.4\textheight]{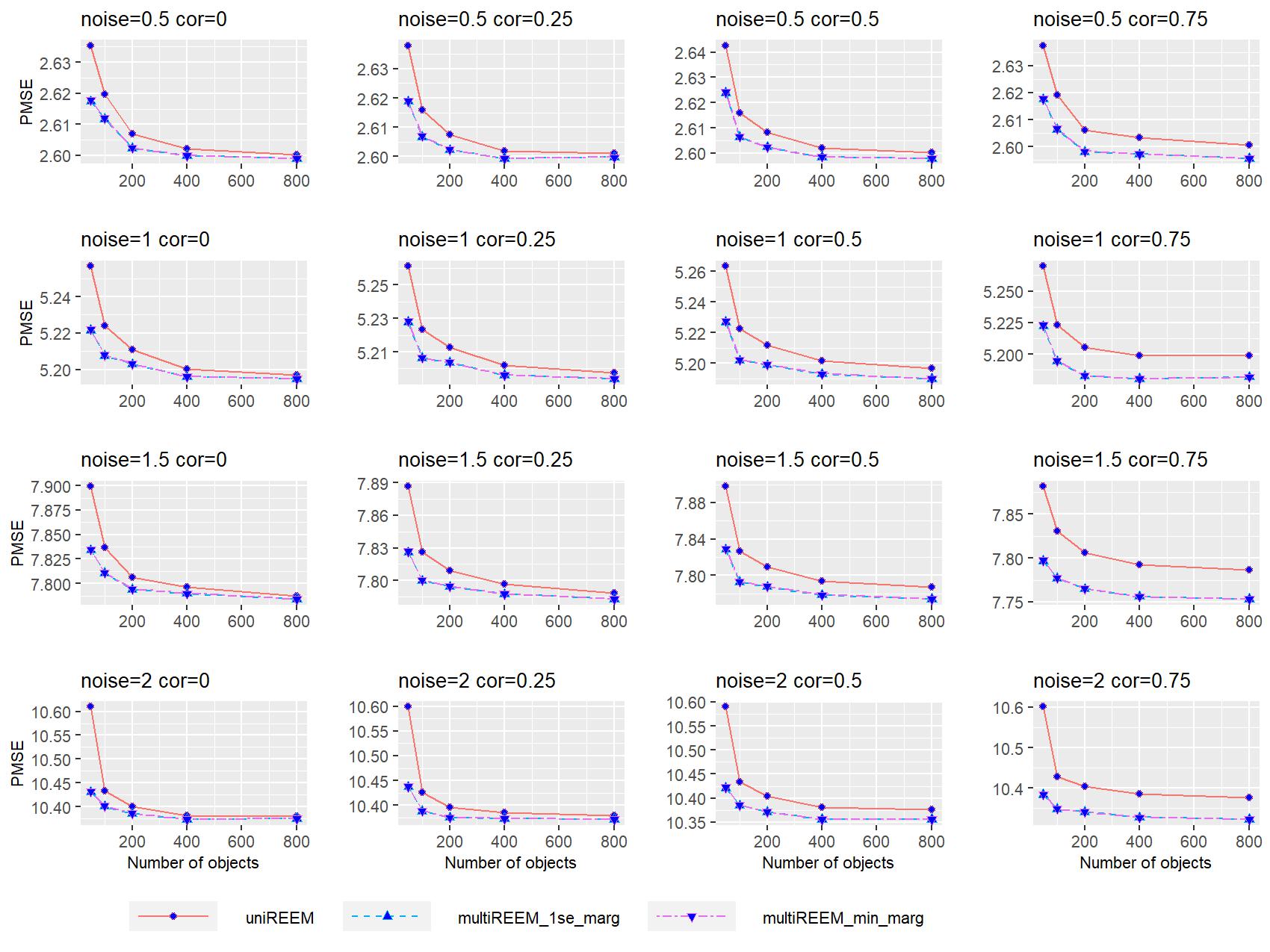}
	\caption{PMSE in object for $T_i=25$ multivariate tree with five response variables}
	\label{fig:ti25-str3-group-part1}
\end{figure}

\begin{figure}[!htp]
	\centering
	\includegraphics[width=1.0\linewidth, height=0.4\textheight]{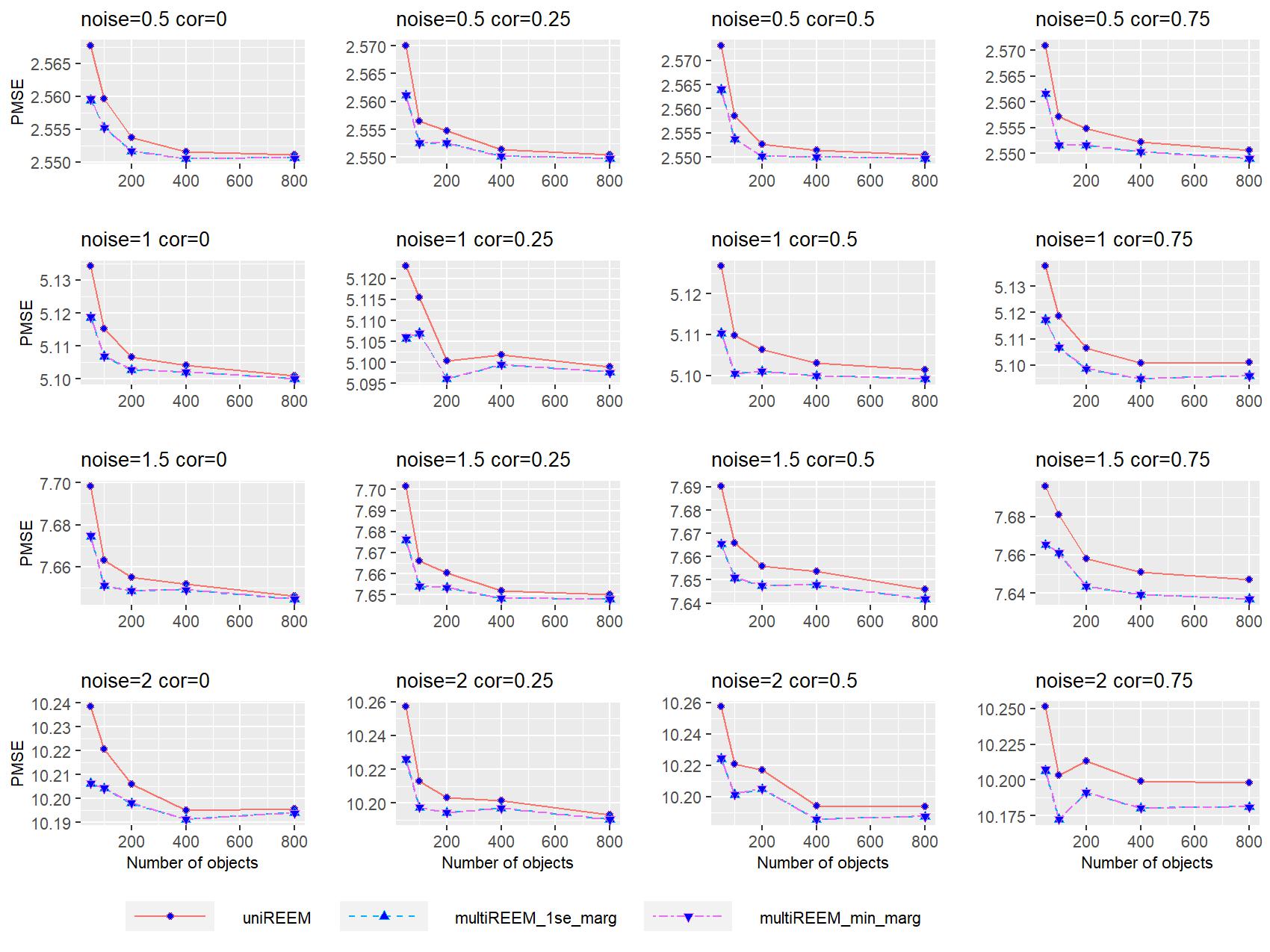}
	\caption{PMSE in object for $T_i=50$ multivariate tree with five response variables}
	\label{fig:ti50-str3-group-part1}
\end{figure}

\begin{figure}[!htp]
	\centering
	\includegraphics[width=1.0\linewidth, height=0.4\textheight]{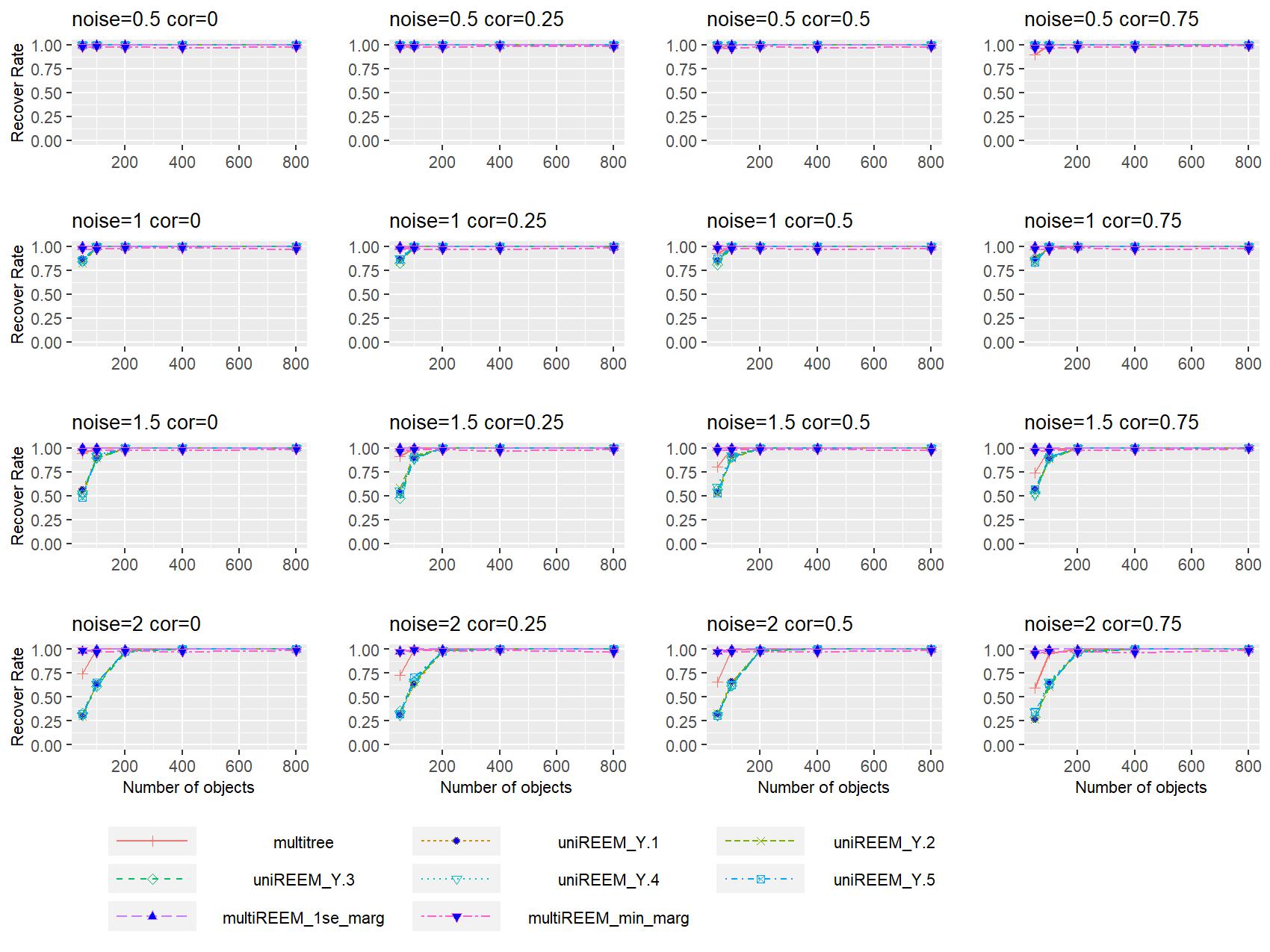}
	\caption{Recovering rate for $T_i=10$ multivariate tree with five response variables}
	\label{fig:ti10-str3-rec}
\end{figure}

\begin{figure}[!htp]
	\centering
	\includegraphics[width=1.0\linewidth, height=0.4\textheight]{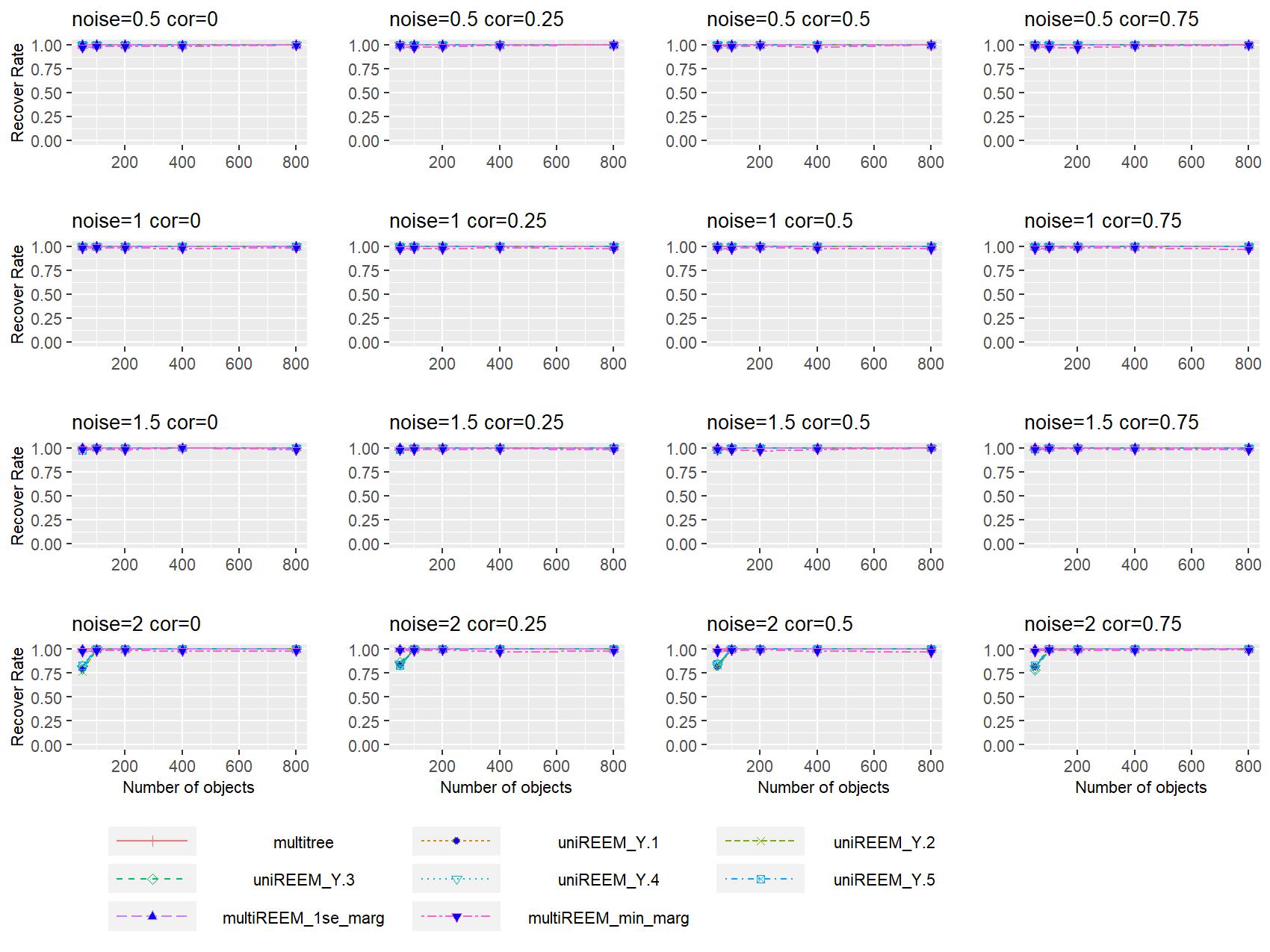}
	\caption{Recovering rate for $T_i=25$ multivariate tree with five response variables}
	\label{fig:ti25-str3-rec}
\end{figure}

\begin{figure}[!htp]
	\centering
	\includegraphics[width=1.0\linewidth, height=0.4\textheight]{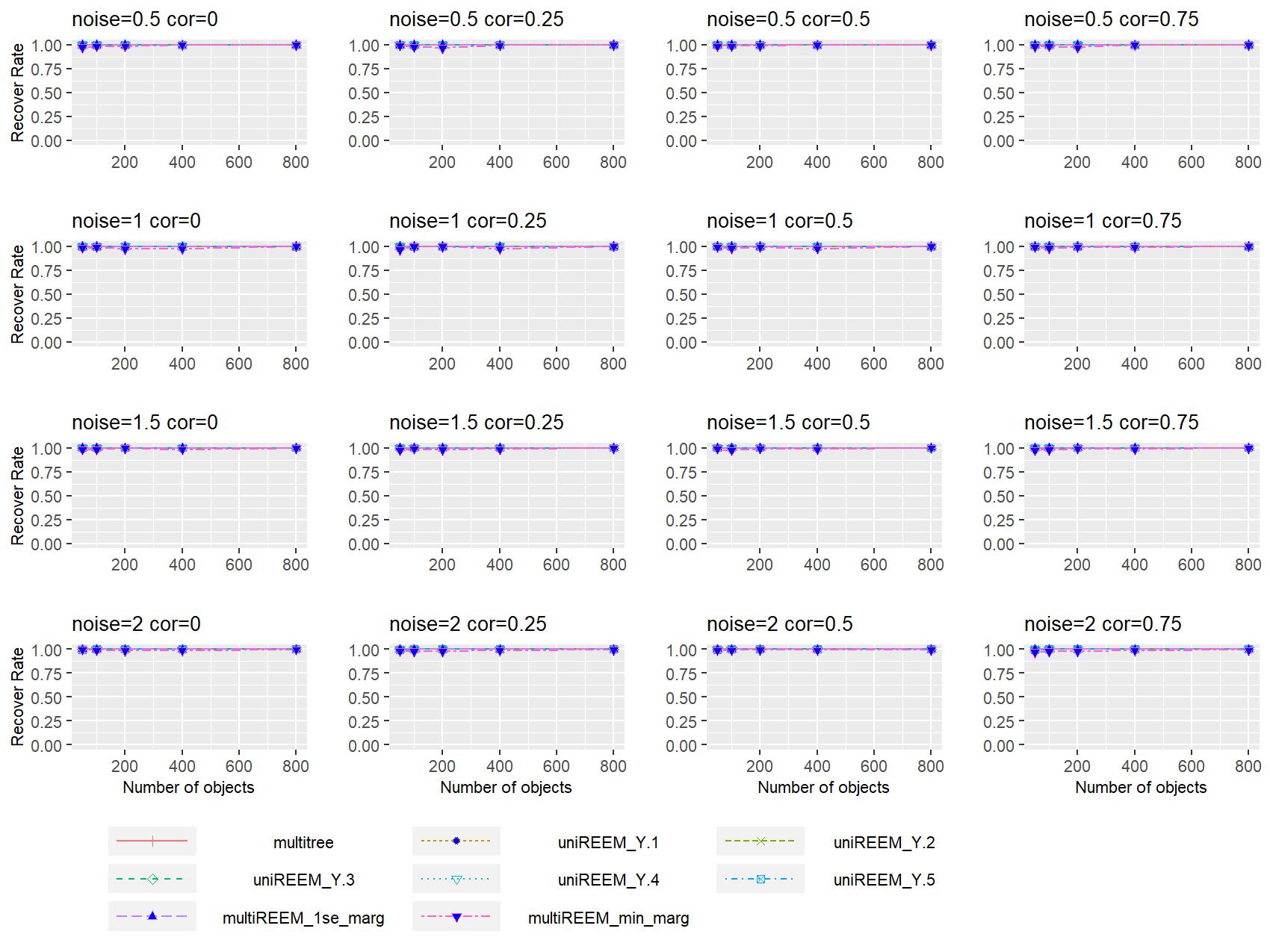}
	\caption{Recovering rate for $T_i=50$ multivariate tree with five response variables}
	\label{fig:ti50-str3-rec}
\end{figure}

The PMSEs and the tree recovering rates for $T_i=10, 25, 50$ with the five-response tree structure are shown in Figures \ref{fig:ti10-str3-group-part1}--\ref{fig:ti50-str3-rec}.



\end{document}